\documentclass[10pt,journal]{IEEEtran}

\usepackage{cite}
\usepackage{xspace}
\usepackage{amsmath,amssymb,amsfonts}
\usepackage[breakable, skins]{tcolorbox}
\usepackage{algorithmic}
\usepackage{graphicx}
\usepackage{textcomp}
\usepackage{xcolor}
\usepackage{caption}
\usepackage{subcaption}
\usepackage{multirow}
\usepackage{colortbl}
\usepackage{booktabs}
\usepackage{color, soul}
\usepackage{hhline}
\usepackage{tabularx}
\usepackage{adjustbox}
\usepackage[hyphens]{url}
\usepackage{enumitem}
\usepackage{balance}
\usepackage{dblfloatfix}
\usepackage[absolute,showboxes]{textpos}
\usepackage[normalem]{ulem}

\tcbset{arc=0mm,size=fbox,attach title to upper={\ ---\ },coltitle=black}

\usepackage{hyperref}
\hypersetup{pdfstartview=FitV,
  breaklinks=true, colorlinks=true, linkcolor=black, citecolor=black, urlcolor=blue,
  pdftitle={A Longitudinal View at the Adoption of Multipath TCP}, pdfauthor={}, pdfkeywords={}
}

\usepackage{cleveref}
\crefformat{section}{\S#2#1#3}
\crefformat{subsection}{\S#2#1#3}
\crefformat{subsubsection}{\S#2#1#3}
\crefrangeformat{section}{\S\S#3#1#4 to~#5#2#6}
\crefmultiformat{section}{\S\S#2#1#3}{ and~#2#1#3}{, #2#1#3}{ and~#2#1#3}

\usepackage{todonotes} %
\presetkeys{todonotes}{inline,color=olive!30}{}

\newcommand{\zmap}{ZMap\xspace}
\newcommand{\tracebox}{Tracebox\xspace}

\newcommand{\eg}{e.g.,\xspace}
\newcommand{\ie}{i.e.,\xspace}
\newcommand{\etal}{\textit{et al.}\xspace}
\newcommand{\one}{(1)\xspace}
\newcommand{\two}{(2)\xspace}
\newcommand{\three}{(3)\xspace}
\newcommand{\four}{(4)\xspace}

\begin{document}
\bstctlcite{IEEEexample:BSTcontrol}

\title{A Longitudinal View at the \\ Adoption of Multipath TCP}

\author{Tanya~Shreedhar,
		Danesh~Zeynali,
        Oliver~Gasser,
        Nitinder~Mohan,
        and~Jörg~Ott}%

\maketitle

\begin{abstract}

Multipath TCP (MPTCP) extends traditional TCP to enable simultaneous use of multiple connection endpoints at the source and destination. MPTCP has been under active development since its standardization in 2013, and more recently in February 2020, MPTCP was upstreamed to the Linux kernel.
In this paper, we provide an in-depth analysis of MPTCPv0 in the Internet and the first analysis of MPTCPv1 to date. We probe the entire IPv4 address space and an IPv6 hitlist to detect MPTCP-enabled systems operational on port 80 and 443. Our scans reveal a steady increase in MPTCPv0-capable IPs, reaching 13k+ on IPv4 (2$\times$ increase in one year) and 1k on IPv6 (40$\times$ increase).
MPTCPv1 deployment is comparatively low with $\approx$ 100 supporting hosts in IPv4 and IPv6, most of which belong to Apple.
We also discover a substantial share of seemingly MPTCP-capable hosts, an artifact of middleboxes mirroring TCP options. We conduct targeted HTTP(S) measurements towards select hosts and find that middleboxes can aggressively impact the perceived quality of applications utilizing MPTCP.
Finally, we analyze two complementary traffic traces from CAIDA and MAWI to shed light on the real-world usage of MPTCP. We find that while MPTCP usage has increased by a factor of 20 over the past few years, its traffic share is still quite low.

\end{abstract}

\section{Introduction}\label{sec:introduction}

Despite significant advances in Internet infrastructure and connectivity, TCP's connectivity model has remained largely unchanged over the last 30 years. 
Recent advances in network technologies have led to the rise of multi-homed devices, \eg smartphones, with access to more than one networking interface.
Multipath TCP (MPTCP) is an extension to TCP that allows endpoints to simultaneously utilize multiple interfaces 
for concurrent or backup data transmissions~\cite{overviewMPTCP}.
Standardized in early 2013, MPTCP has shown better resource utilization, higher aggregated throughput, and resilience to network failures in numerous research studies published over the years~\cite{shreedhar2018qaware, mptcp-edge, blest, mptcp-cross, mptcp-datacenters}.

Due to the performance benefits of MPTCP compared to TCP, several known organizations have incorporated the protocol within their products and services. 
Apple uses MPTCP in its iOS devices to enhance the user experience surrounding its system services, \eg 
Siri, Music, Maps, Wi-Fi Assist~\cite{apple_backup}.
In 2019, Apple provided APIs to third-party developers for making use of MPTCP in non-system iOS applications.
Korea Telecom, in partnership with Samsung, uses MPTCP to provide Gigabit speeds over Wi-Fi and LTE to its customers~\cite{kt-gigalte}.
In February 2020, MPTCPv1 was upstreamed to Linux and is now available to all users running Linux 5.6 or newer~\cite{upstream_linux}. 

Despite significant interest in improving the protocol~\cite{overviewMPTCP,paasch2014scheduler,shreedhar2018qaware}
, the current state of MPTCP deployment in the Internet remains largely unexplored in research.
We attribute this gap partially to the influence of middleboxes on the accuracy of such studies.
The Internet is proliferated with a wide spectrum of specialized appliances and systems known as middleboxes that meddle with user traffic before it reaches the target~\cite{sherry2012making}. 
The intended operation of middleboxes is to offer valuable benefits, \eg firewalls drop unintentional packets and proxies improve the performance of connection setup.
However, certain middleboxes interact quite poorly with connections containing TCP header extensions.
While some may strip the packet 
of any header additions before relaying it to the next hop, others might block the connection altogether~\cite{hesmans2013tcp}.
Since MPTCP relies on TCP extensions for signaling, it is also susceptible to such middleboxes in the Internet.
MPTCP designers incorporate several mechanisms into the protocol specification that allows the protocol to fall back to regular TCP
for data transfers, if the connection is affected by middleboxes~\cite{rfc8684}.
Despite that, middleboxes continue to hinder MPTCP studies, since %
scanning tools leverage the connection establishment mechanism to interact with targets and thus remain vulnerable to side-effects of middleboxes.    
In a study from 2015, the authors wanted to analyze the deployment of MPTCP in the Internet, it later became clear that the results include false-positives due to middleboxes echoing MPTCP options for non-MPTCP hosts~\cite{early_look}.
However, despite significant measurement challenges, assessing the adoption of MPTCP in-the-wild is still pertinent since the protocol can only be employed if there is sufficient server-side support in the Internet.

This paper extends our previous study \cite{aschenbrennersingle} and presents a broad and multi-faceted assessment of MPTCPv0 and MPTCPv1.
We study both the \textit{infrastructure}, in terms of MPTCP-capable IPv4 and IPv6 addresses, and the \textit{traffic share} at two geographically diverse vantage points.
We identify and remove middleboxes affecting MPTCP operation in-the-wild from our scans to avoid false-positives. 
Furthermore, we also investigate if such middleboxes also negatively impact MPTCP application traffic.
Specifically, we make the following contributions in our paper. 

\begin{enumerate}[label=\Roman*., nolistsep,labelwidth=!, labelindent=0pt]
    \item We regularly probe the entire IPv4 address space and an IPv6 hitlist~\cite{gasser2018clusters} for MPTCPv0 support since July 2020 using \zmap. 
Our scans target HTTP (port 80) and HTTPS (port 443) since they make up the largest traffic share in the Internet~\cite{trevisan2020five,feldmann2020lockdown}.
We find that our scans are affected mainly by middleboxes that echo TCP extensions, indicating that traditional scanning methods are still ineffective in accurately evaluating the deployment of MPTCP.
We also observe that the number of IPs reported to support MPTCP \emph{without} replayed options increased fourfold over IPv4 port 443 compared to 2015 \cite{early_look}.

    \item We present the first analysis of the MPTCPv1 ecosystem to date and regularly perform measurements since March 2021 on ports TCP/80 and TCP/443.
        Its support on both IPv4 and IPv6 remains almost non-existent before October 2021 and remains low compared to MPTCPv0 afterwards.
        In a case-study, we extend our measurement period to February 2022, and
        we find that Apple has added MPTCPv1 support to some of its hosts and is now dominating its deployment in IPv4 and IPv6.

    \item We scrutinize targets that reportedly support MPTCP in our \zmap scans for middleboxes by using \tracebox~\cite{tracebox} and make two significant discoveries. 
First, some MPTCP-capable hosts in the IPv4 are \emph{transient}; indicating experimental connotations attached to MPTCP usage in-the-wild.
Second, we identify several middleboxes that interact in a much more complicated fashion than just echoing with MPTCP packets.
Despite that, we observe a growing adoption of MPTCP reaching 16.5k/13.5k and 1195/1184 over port 80/443 on IPv4 and IPv6, respectively.
Compared to December 2020, MPTCP support has grown substantially: 2$\times$ in IPv4 and 40$\times$ in IPv6.
 
    \item We initiate parallel HTTP(S) GET requests using MPTCP and regular TCP towards IPs identified in our \zmap and \tracebox measurements.
Our results show that a majority of \emph{truly} MPTCP-capable servers are indifferent to the choice of a transport protocol for connection establishment.
However, IP addresses affected by middleboxes take longer to successfully establish a connection using MPTCP vs. TCP, hinting at the potential impact of middleboxes on MPTCP's perceived quality.

    \item We analyze the usage of MPTCP for data transfers over the Internet by investigating four years of inter-domain traffic collected by CAIDA and MAWI.
Our findings show that MPTCP data usage is still quite low compared to TCP (peaking at 0.4\%), primarily due to the lack of widespread MPTCP support among clients, servers, and applications.
We find that Apple---a vocal supporter of MPTCP---is responsible for almost all MPTCP traffic that we observe.
Moreover, in early 2022 we can find the first data being exchanged over MPTCPv1 in-the-wild.
Finally, we observe a steadily rising popularity of MPTCP for large data transfers.
\end{enumerate}

\noindent To foster reproducibility, we plan to publish our latest datasets and scripts similar to the previous iteration of this work \cite{mptcpDataSet}. 
We also continuously perform MPTCPv0 and MPTCPv1 scans and publish the results at \texttt{\url{https://mptcp.io}}.
\section{Multipath TCP: An Overview}\label{sec:background}

Multipath  TCP  (MPTCP), standardized by the IETF in 2013, is a multipath extension to TCP.
It utilizes multiple interfaces of a given device to create multiple subflows and transmits data using these multiple subflows concurrently. This leads to better resource utilization, higher aggregated throughput, and resilience to network failures \cite{rfc8684,overviewMPTCP}. 
The subflows can be added and removed throughout the MPTCP lifecycle. 
However, both end-hosts must be MPTCP enabled to support multipath transfer.
Data from the unmodified applications are scheduled on one of the underlying TCP subflows by the MPTCP scheduler block based on the scheduling policy~\cite{paasch2014scheduler}.
The default \texttt{minSRTT} scheduler~\cite{minsrtt} sends data on the available subflow with the lowest smoothed round trip time (SRTT) to the receiver. 
MPTCP's \emph{coupled congestion control} mechanisms balance packet congestion over all the subflows and minimize additional reordering delays~\cite{raiciu2012}.

\begin{figure}[t!]
	\centering
	\includegraphics[width=\columnwidth]{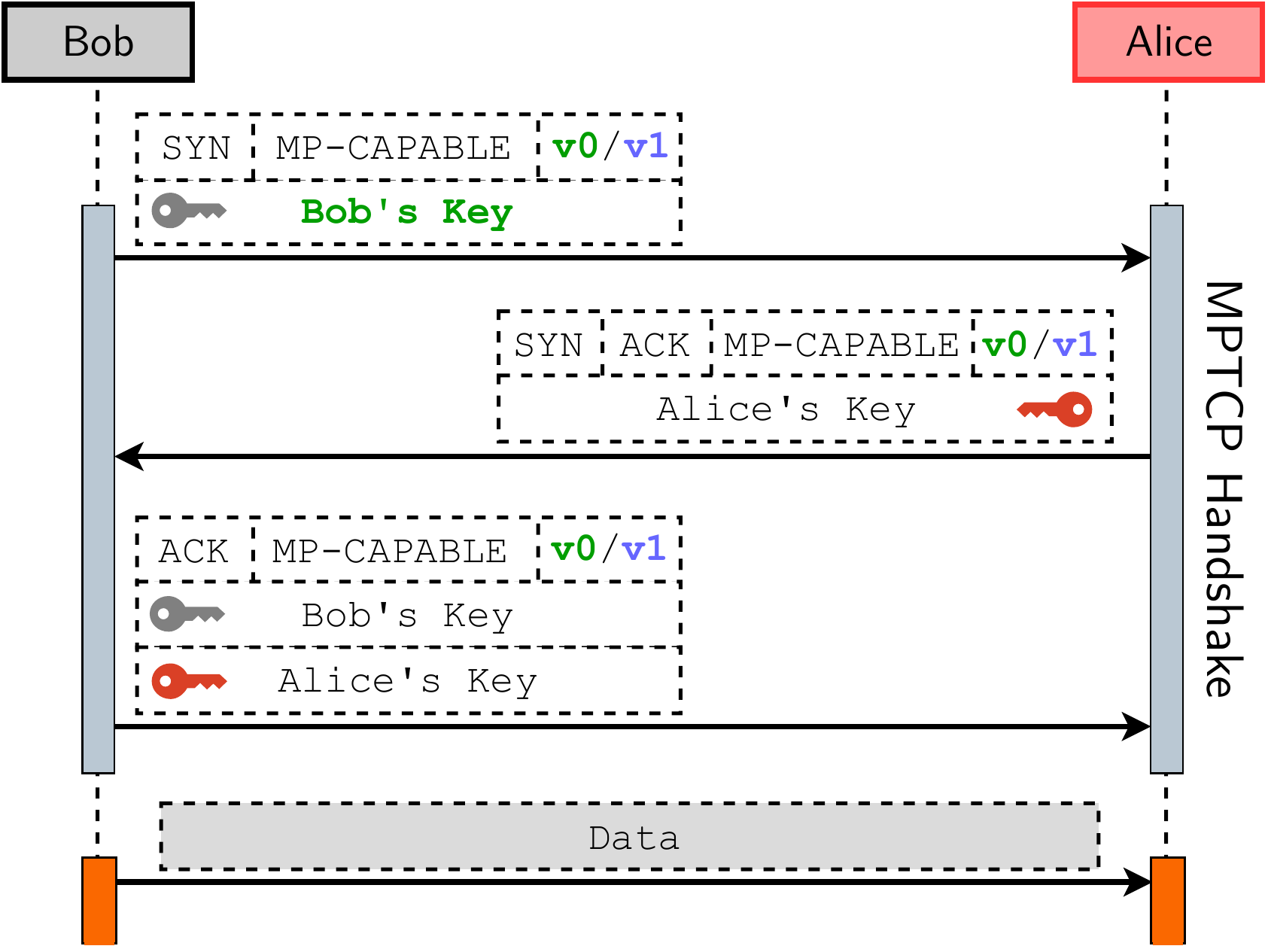}
	\caption{MPTCPv0 and MPTCPv1 connection establishment process between two MPTCP-capable endpoints~\cite{rfc6824,rfc8684}. The \texttt{MP\_CAPABLE} option is slightly different for the both versions. While MPTCPv0 SYN includes Bob's sender's key (shown in green dashed box), MPTCPv1 SYN is sent without the key and Bob's key is sent in the final ACK.}
	\label{fig:key_exchange}
\end{figure}

We focus here on the MPTCP connection establishment procedure
as our measurement approach utilizes it inherently.
We refer inclined readers to previous work~\cite{overviewMPTCP,paasch2014scheduler} for more details on MPTCP machinery, features, and design choices.
\Cref{fig:key_exchange} shows the MPTCPv0 and MPTCPv1 connection establishment process between MPTCP-enabled client and server.
The MPTCP handshake mechanism is derived from the TCP three-way handshake.
In addition, MPTCP hosts use a random 64-bit sequence as \emph{keys} to authenticate themselves when setting up new subflows~\cite{rfc6824,rfc8684}.
Moreover, every packet in the handshake signals MPTCP support through the \texttt{MP{\_}CAPABLE} option.
The \texttt{MP{\_}CAPABLE} option also includes the version supported by the host.
The connection establishment process differs for MPTCPv0 and MPTCPv1.
For MPTCPv0, the client (in our example Bob) initiates a connection by sending a SYN packet containing its key (highlighted by green) and the \texttt{MP{\_}CAPABLE} option
to the server (Alice).
For MPTCPv1, Bob's sender's key is not sent in the SYN packet.
In case the host supports multiple versions, the highest version number is sent in the \texttt{MP{\_}CAPABLE} option.
If the server also supports MPTCP, it replies back with a SYN-ACK including the \texttt{MP{\_}CAPABLE} option and its own key.
The receiver's \texttt{MP{\_}CAPABLE} option will indicate the version number it supports.
According to the specification~\cite{rfc6824,rfc8684}, both key values in SYN and SYN-ACK are individually referred to as Bob's and Alice's sender's key.
In the first stage of our study, we use \texttt{MP{\_}CAPABLE} SYN packets to probe for hosts which reply with an \texttt{MP{\_}CAPABLE} option in the SYN-ACK; recording their IP address and sender's key value (for more details see \Cref{sec:zmapMethod}).
For both MPTCPv0 and MPTCPv1, Bob finally establishes the connection by sending an ACK with both keys and the \texttt{MP{\_}CAPABLE} option.
This allows regular MPTCP data transmissions between the two parties.

\section{Related Work} \label{sec:related}

\begin{figure*}[t!]
	\begin{subfigure}[t]{\textwidth}
		\centering
		\includegraphics[width=\textwidth]{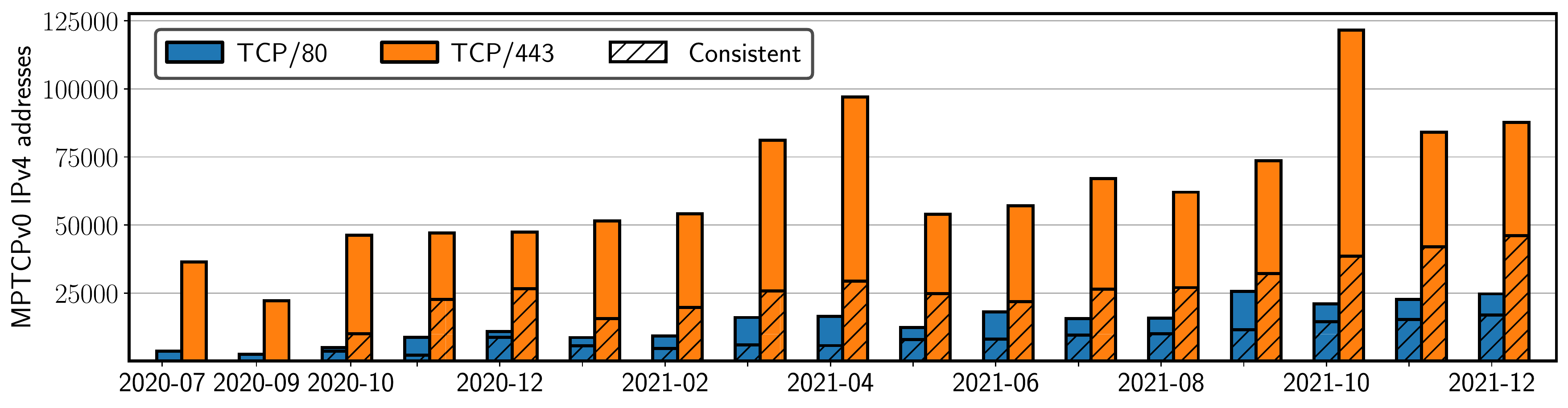}
		\caption{IPv4 addresses in our scans from July 2020 -- December 2021.}
		\label{fig:zmap-v0-mpdiff1}
	\end{subfigure}
	\vspace*{0.3em}
	\begin{subfigure}[t]{\textwidth}
		\centering
		\includegraphics[width=\textwidth]{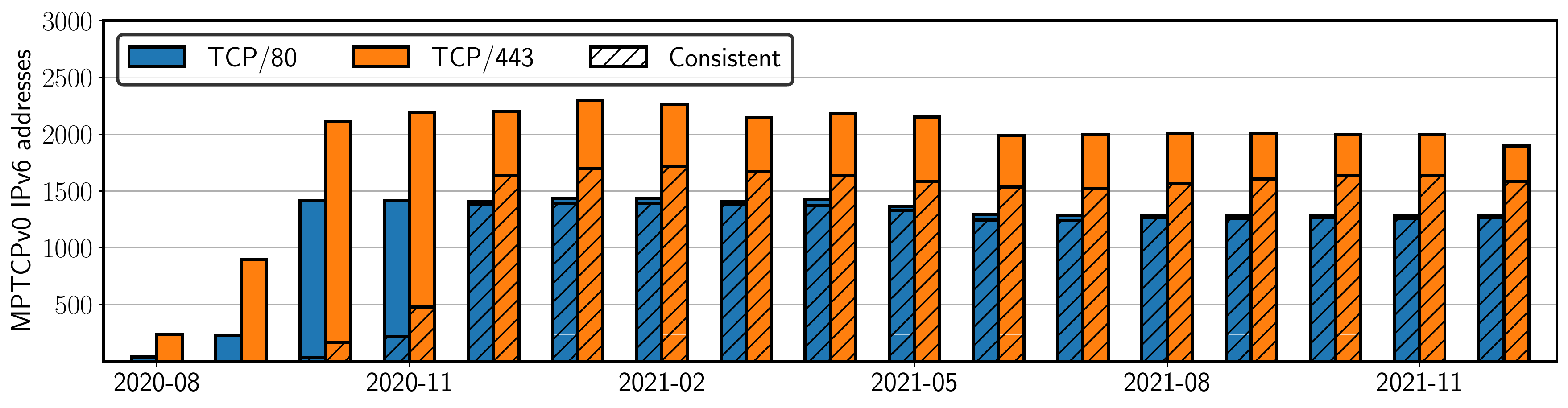}
		\caption{IPv6 addresses in our scans from August 2020 -- December 2021.
		}
		\label{fig:zmap-ipv6-v0}
	\end{subfigure}
	\caption{Unique (a) IPv4 and (b) IPv6 addresses in our \zmap scans for MPTCPv0 over port 80 and port 443 over 18-months that returned a different MPTCP key than the one sent in SYN. The ``consistent'' counts the common hosts that are consistently available across three month moving average window.}
    \label{fig:mptcpv0-zmap}
\end{figure*}

Scanning the activity of different protocols in the Internet has been a long-lasting interest within the network measurement research community~\cite{dai2021smap,izhikevich2021lzr}.
In early 2008, Heidemann \etal~\cite{heidemann2008census} systematically probed a subset of 1\% of IPv4 address space with ICMP pings.
The state of active scanning research was pushed forward significantly by \zmap \cite{durumeric2013zmap}, which allows researchers to scan the entire IPv4 address space in less than an hour.
Several works have since used the tool to investigate the deployment of different protocols and applications in the Internet, \eg liveness~\cite{bano2018scanning}, TCP initial window~\cite{conf/imc/ruth2017}, and QUIC~\cite{conf/pam/jan2018}.
Others have looked into passive data traces for a different viewpoint on deployment measurements.
Richter \etal~\cite{conf/imc/richter2016} studied the IPv4 activity as observed from within the Akamai network and show that comparative active scanning studies miss up to 40\% of the hosts that contact the CDN.
Qian \etal~\cite{conf/sigcomm/qian2009} analyzed TCP behavior
from multiple vantage points within a large tier-1 ISP.
Maghsoudlou \etal~\cite{maghsoudlou2021zeroing} used active and passive scanning to analyze and characterize the traffic from port 0.
Wan \etal~\cite{conf/imc/wan2020} discussed several factors that can impact the quality of scanning results \eg geo-location, losses, blocking, etc.
We circumvent these biases to the best of our abilities by following best practices and scanning continuously for six months.
Please refer to \Cref{sec:zmapMethod} for our detailed measurement methodology. 

\smallskip
\noindent \textbf{Transport Protocol Deployment in the Internet.}
There is a substantial body of work that has looked into the deployment of transport layer protocols over the Internet, the majority of which has been highly focused on QUIC~\cite{quic}.
QUIC is a recently standardized transport protocol that has seen widespread deployment~\cite{rfc9000}. Rüth \etal~\cite{conf/pam/jan2018} performed ZMap scans over IPv4 to check for QUIC support in the wild. They further built a scanning tool using \texttt{quic-go}\footnote{https://github.com/lucas-clemente/quic-go}
to investigate QUIC. The results indicate that QUIC's traffic over the Internet is continuously increasing and upwards of $\approx 8\%$ with $\approx 98\%$ of that traffic going to/from the Google AS~\cite{conf/pam/jan2018}. A more recent study~\cite{zirngibl2021quicDeployment} checks for the deployment of different QUIC versions by developing ZMap modules for both IPv4 and IPv6. They also developed \texttt{QScanner}, a scanning tool to analyze the various deployments in detail. Madariaga \etal~\cite{madariaga2020adoption} analyze the adoption of QUIC from measurements in user-space taken by mobile end-user devices. The authors developed an Android framework to perform network flow measurements through passive monitoring of active connections.
There are works that have measured other important aspects of the protocol, such as DNS over QUIC~\cite{kosek2022dnsOverQuic} and web censorship measurements of HTTP/3 over QUIC~\cite{elmenhorst2021webCensorship}.

The closest work to ours dates back to 2015~\cite{early_look}.
Mehani \etal proposed a scanning mechanism that probed every host on port 80 of the Alexa Top 1M with \zmap
and classified IP addresses that responded with \texttt{MP\_CAPABLE} as supporting MPTCP.
Their results indicate that
less than 0.1\% of scanned targets support MPTCP, with a majority located in China.
However, the accuracy of the work was later found to be low as it falsely recognized
middleboxes that echoed unknown TCP extensions as MPTCP hosts~\cite{not_easy}. The authors later published an errata and tracked the non middlebox-affected MPTCP deployment for several months in 2015.
In a previous iteration of this work~\cite{aschenbrennersingle}, the authors extend the above methodology to identify middleboxes affecting MPTCP correctly, hence providing the most accurate picture of \emph{true} MPTCPv0 deployment.
The authors observed a steady growth in MPTCP-enabled IPs that support HTTP and HTTPS, primarily driven by Apple.
This work provides the first long-term multi-faceted analysis of MPTCP v0 and MPTCP v1 deployment in the Internet. Additionally, we identify the \emph{true} support for MPTCP over the two most popular services in the Internet, HTTP and HTTPS, over both IPv4 and IPv6.
\section{Active Internet Scans}\label{sec:activescans}

To identify support for Multipath TCP in the Internet, we actively scan for MPTCP options (both \textit{version 0} and \textit{version 1}) over the IPv4 and IPv6 address space.
Our study probes the entire IPv4 address space over port 80 ($
	\approx$ 74M unique responsive IPs) and port 443 ($\approx$ 52M unique responsive IPs).
In IPv6, we use the \emph{IPv6 hitlist}~\cite{gasser2018clusters} to probe both port 80 and port 443 due to the size of the address space.
We find 746k and 544k responsive IPv4 and IPv6 addresses, respectively.
Our results in this work extend our previous measurement study~\cite{aschenbrennersingle} and are drawn from over 18 months of data collection (July 2020 -- December 2021).
We share our dataset and results on the website: \texttt{\url{https://mptcp.io}}

\subsection{Methodology}\label{sec:zmapMethod}
We use \zmap~\cite{zmapv6} to rapidly enumerate IPv4 and IPv6 addresses.
To identify MPTCP hosts (both MPTCPv0 and MPTCPv1), we leverage the initial handshake mechanism.
However, since the MPTCP options sent during the connection establishment procedure differ for both versions (as discussed in \Cref{sec:related}), we perform separate periodic scans for each version.
As shown in \Cref{fig:key_exchange}, in MPTCPv0, we send a SYN with the \texttt{MP\_CAPABLE} option for version number 0 along with a static sender's key.
On the other hand, for MPTCPv1, we send the SYN with the \texttt{MP\_CAPABLE} option for version number 1 only (i.e., we do not send a static sender's key in our MPTCPv1 scans).
We record the SYN-ACK responses for scans of both versions.
As illustrated in \Cref{fig:key_exchange}, a legitimate MPTCP host (both versions) will reply back to an MPTCP SYN with a SYN-ACK containing the \texttt{MP\_CAPABLE} option, version number and its own sender's key.
If the target's SYN-ACK response includes these values, we classify it as \emph{potentially MPTCP-capable} hosts.
Previous research has shown that the accuracy of identifying MPTCP hosts via \zmap can be low due to middleboxes that replay or strip packets with TCP extensions~\cite{early_look, not_easy, hesmans2013tcp}.
As shown in previous work~\cite{aschenbrennersingle}, such middleboxes particularly affect MPTCPv0 as the sender's key value is repeated back to the host---resulting in a large number of false positives.
We discuss and highlight this issue further in \Cref{sec:tracebox}.
Such ill-interaction between middleboxes and MPTCPv0 operation primarily motivated MPTCP protocol designers to remove the sender's key from SYN during connection establishment~\cite{rfc8684}.
To further improve the correctness of our analysis for MPTCPv0, we probe \emph{potentially MPTCP-capable} hosts with the well-known middlebox-detection tool \tracebox~\cite{tracebox}.
\tracebox allows us to detect the presence of MPTCP options modifications on the path, revealing
IPs that \emph{truly} support MPTCP.
For our MPTCPv1 scans, we discard SYN-ACK responses that simply mirror our MPTCP option sent in the SYN.
For both MPTCPv0 and MPTCPv1, we filter out responses that do not include the expected version number in MPTCP options.

Before conducting active measurements, we incorporate proposals by Partridge and Allman~\cite{partridge2016ethical} and Dittrich \etal \cite{dittrich2012menlo}.
We follow best scanning practices~\cite{durumeric2013zmap} by limiting our probing rate, maintaining a blocklist, and using dedicated servers with informing rDNS names, websites, and abuse contacts.
Furthermore, we diligently complied to any emails from organizations asking for their networks to be blocklisted.

\begin{figure*}[t!]
	\begin{subfigure}[t!]{\columnwidth}
	\centering
	\includegraphics[width=\columnwidth]{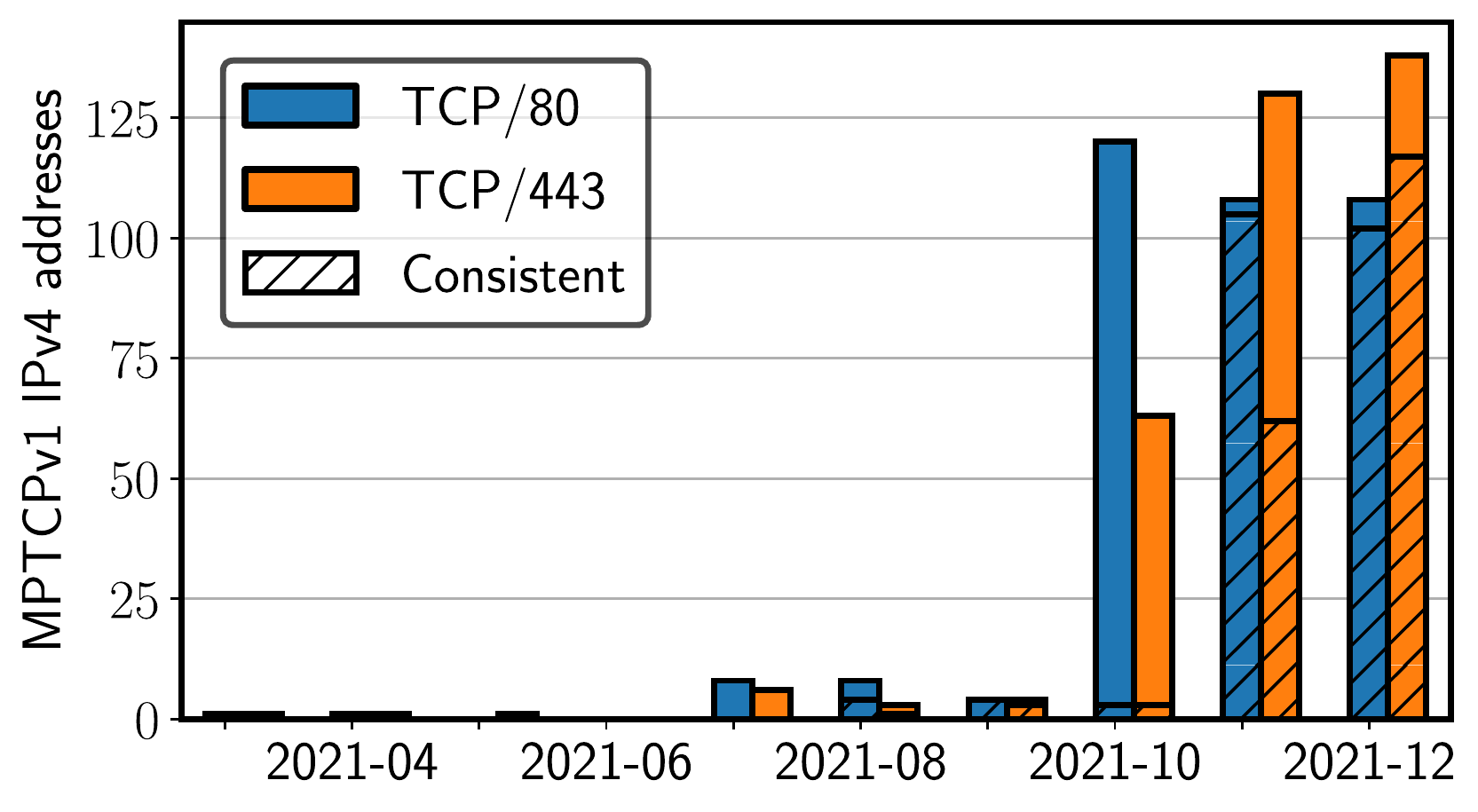}
	\caption{IPv4 addresses in our scans from March -- December 2021
	}
	\label{fig:zmap-v1}
\end{subfigure}
\begin{subfigure}[t!]{\columnwidth}
	\centering
	\includegraphics[width=\columnwidth]{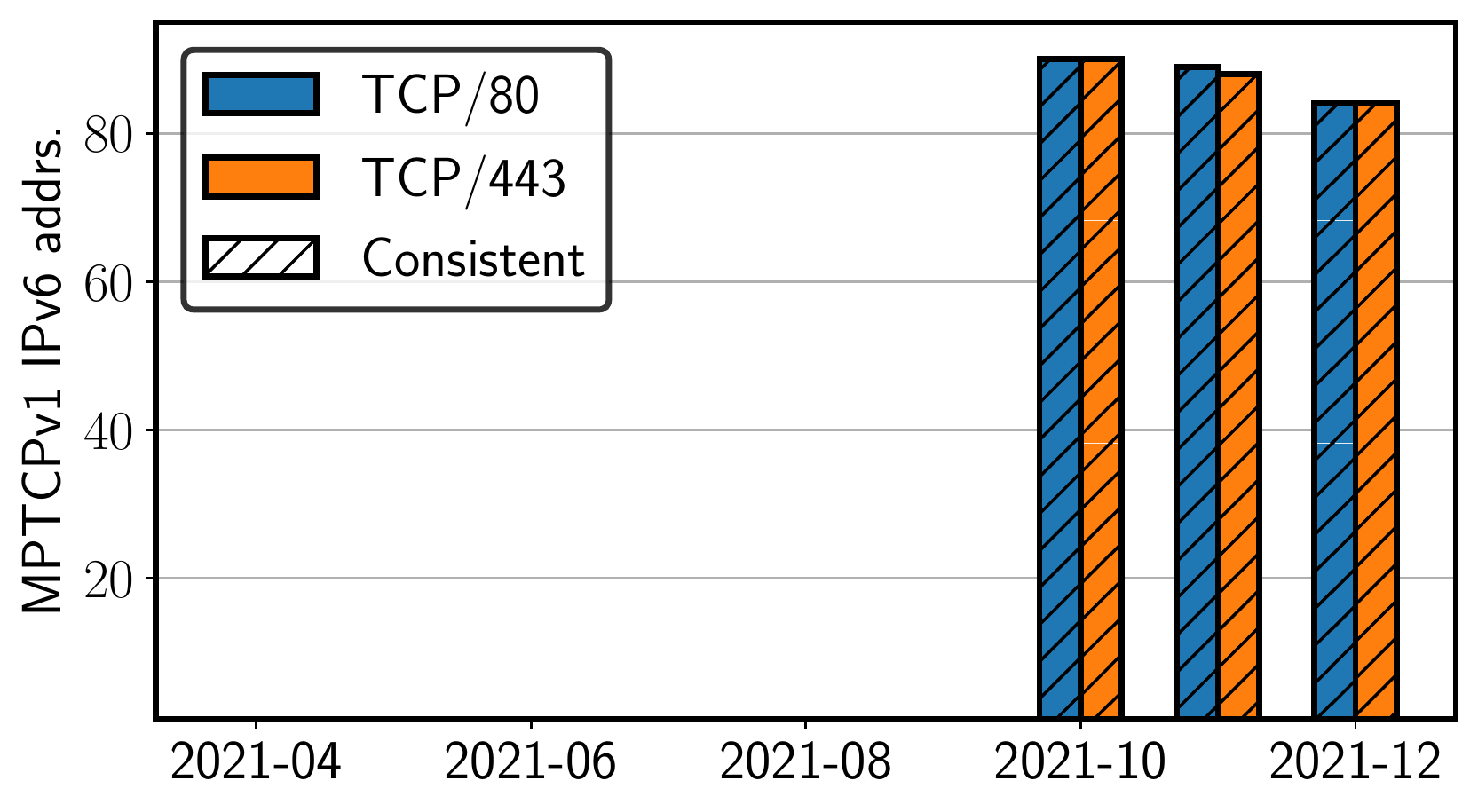}
	\caption{IPv6 addresses in our scans from April -- December 2021
		}
	\label{fig:zmap-ipv6-v1}
\end{subfigure}
\caption{Unique (a) IPv4 and (b) IPv6 addresses in our \zmap scans for MPTCPv1 port 80/443 that returned MPTCP key in response. The ``consistent'' counts the common hosts that are consistently available across three month moving average window.
	}
	\label{fig:mptcpv1-zmap}
\end{figure*}

\subsection{Finding MPTCP Support In-The-Wild} \label{sec:mptcp-zmap}

\subsubsection{ZMap Scan Results}

Our scans targeted almost 60M responsive addresses on port 80 and 50M on port 443 in IPv4. Of these, about 300k addresses for MPTCPv0 and 200k addresses for MPTCPv1 responded with the \texttt{MP\_CAPABLE} flag in their SYN-ACK (\emph{potential MPTCP}) -- making $\approx$5.0\% and $\approx$6.0\% of total responsive hosts on port 80 and 443 for MPTCPv0 and $\approx$3.3\% and $\approx$4.6\% of total responsive hosts on port 80 and 443 for MPTCPv1, respectively. %

\smallskip
\noindent \textbf{MPTCPv0:} Figure~\ref{fig:zmap-v0-mpdiff1} provides a month-wise distribution of our scanning results of MPTCPv0 over the IPv4 address space over port 80 and port 443 for our 18 months scan period from July 2020 to December 2021. 
Note that we were unable to perform any IPv4 port 443 scans in September 2020 due to infrastructural reasons and hence we omit the port 80 results for that month.
Figure~\ref{fig:zmap-ipv6-v0} provides a month-wise distribution of our scanning results of MPTCPv0 over the IPv6 address space over port 80 and port 443 for our 18 months scan period from August 2020 to December 2021.
The bars indicate the unique IP addresses that return a different MPTCP key than the one sent in SYN. The ``consistent'' numbers represent the hosts that are consistently available across the past three-month moving average window. 
We show the ``consistent'' hosts to highlight the stability of MPTCP support over a three-month interval and to isolate any possible effects that we might observe due to transient hosts.
At first glance, it might seem that MPTCPv0 support is increasing consistently during our measurement period, with the highest number of supported addresses reaching $\approx$25k and $\approx$121k on port 80/443 in October 2021, respectively. Moreover, the support for MPTCPv0 on port 443 seems much higher than on port 80 in both IPv4 and IPv6.
However, we also observe that a large percentage of \emph{potential MPTCP} hosts are inconsistently active across our scanning period, signaling at the existence of transient hosts. For IPv4, the consistently available IPs hover around $\approx$50--80\% and $\approx$25--50\% on port 80 and port 443, respectively.
On the other hand, compared to IPv4, we find a very small number of IPv6 addresses responding with the \texttt{MP\_CAPABLE} option. However, the support has increased from our previous observations~\cite{aschenbrennersingle} and remains stable over time since early 2021.

\noindent \textbf{MPTCPv1:} \Cref{fig:zmap-v1,fig:zmap-ipv6-v1} show the unique IPv4 and IPv6 addresses in our \zmap scans for MPTCPv1 for port 80/443 from April -- December 2021 that returned a non-mirrored MPTCPv1 response. To assert MPTCPv1 support, we check the MPTCP version number and filter out all hosts that replay the same MPTCP options as we sent in the SYN packet. Similar to the MPTCPv0 results in \Cref{fig:mptcpv0-zmap}, ``consistent'' count the common IPs that are consistently available across a three-month moving window.

 We observe a large difference in the number of hosts that support MPTCP between v0 and v1. Out of 200K hosts that respond to our scans, MPTCPv1 hosts in both IPv4 and IPv6 hovered close to 100. In fact, we found that MPTCPv1 support was almost non-existent until October 2021, after which we observe a large spike in both IPv4 and IPv6. On closer analysis, we find Apple to be the primary contributor to this support (we explore this trend further in \Cref{sec:discussion:sub:apple}). For months prior to October 2021, MPTCPv1 had minimal support in the Internet, which came mostly from experimental networks. Also, note that the percentage of consistent responsive hosts is much larger in MPTCPv1 compared to MPTCPv0 for both IPv4 and IPv6 -- hovering around 80--90\%.
Finally, we find that the magnitude of responsive MPTCPv1 IPv4 and IPv6 hosts is quite similar ($\approx$ 120 in IPv4 vs. $\approx$ 80 in IPv6), which is in stark contrast to MPCTPv0, where the chasm between the two is much larger ($\approx$ 100k vs. $\approx$ 2k).

\begin{figure}[t]
	\begin{subfigure}{0.5\columnwidth}
		\includegraphics[width=\textwidth]{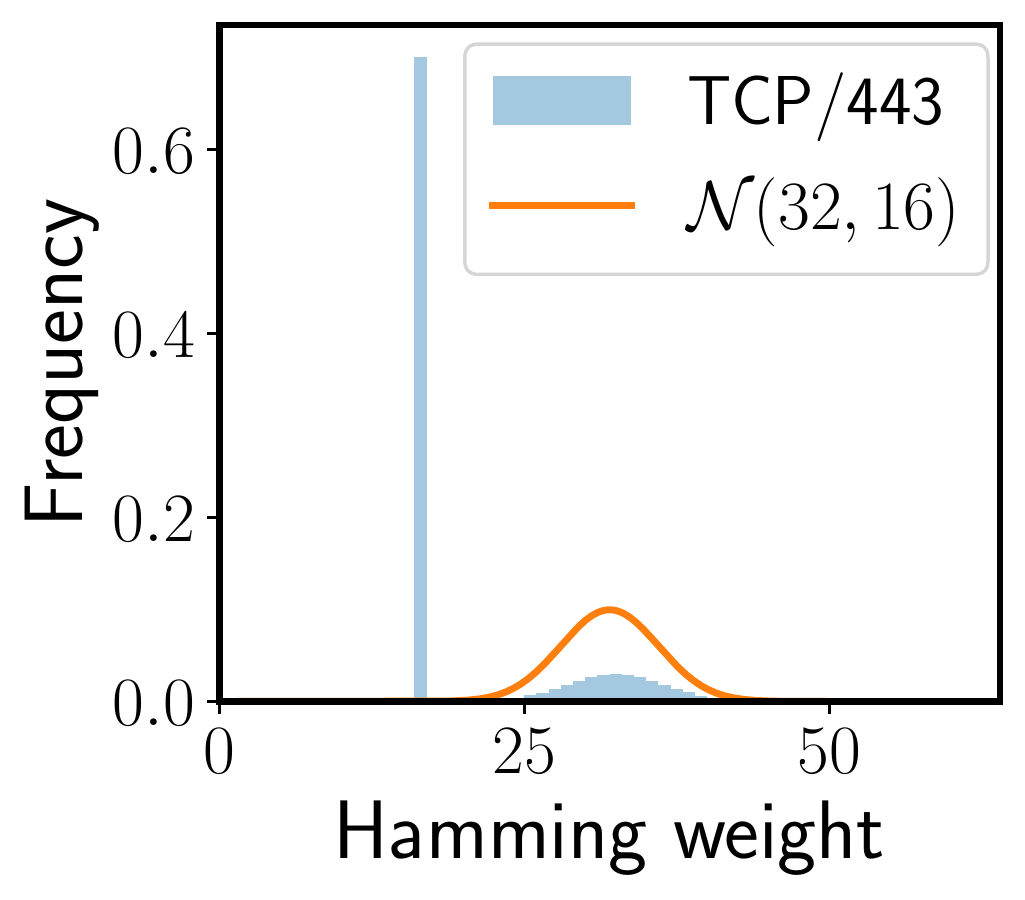}
		\caption{IPv4 Port 443}
		\label{fig:p443v4hamming}
	\end{subfigure}%
	\begin{subfigure}{0.5\columnwidth}
		\includegraphics[width=\textwidth]{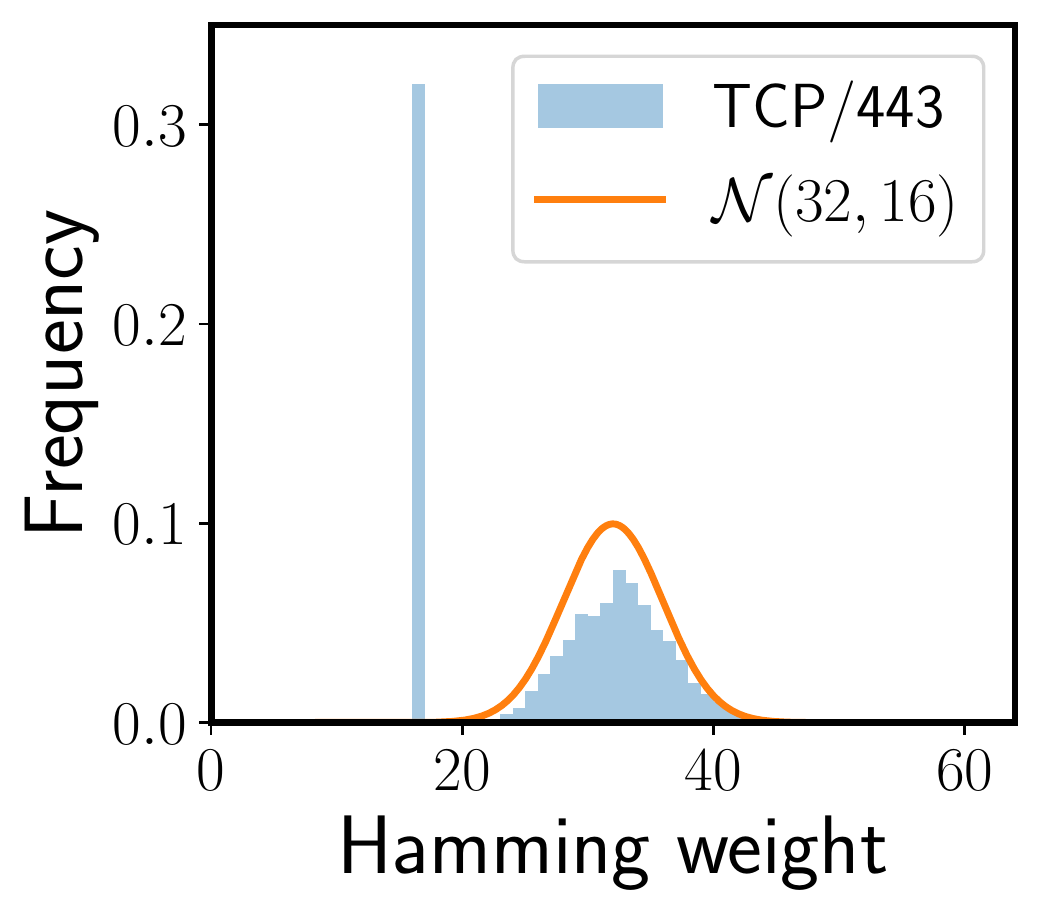}
		\caption{IPv6 Port 443}
		\label{fig:p443v6hamming}
	\end{subfigure}%
	\caption{Hamming weight distribution of sender's keys received from potential MPTCPv0-capable hosts in our \zmap scans.}
	\label{fig:Hamming}
\end{figure}

\smallskip
\subsubsection{Impact of middleboxes on correctness of scans}
To examine if our scans are affected by interfering middleboxes, we analyze the MPTCP sender's key we receive from targets in their SYN-ACK response.
A \emph{true} MPTCP host generates a random 64-bit sequence to use as the key (see \Cref{sec:related}).
According to the central limit theorem, the sum of independent random variables tends toward a normal distribution.
In that case, the sum of all bits in the sender's key---\ie the Hamming weight---should follow the normal distribution $\mathcal{N}(32, 16)$.
\Cref{fig:Hamming} shows the Hamming weight distribution of the sender's key from potential MPTCPv0 hosts on port 443 in IPv4 and IPv6.
We find that a large number of sender's keys do not follow the normal distribution.
In fact, the Hamming weight 16, \ie the exact Hamming weight of the key that we send in our SYN probes, is heavily over-represented.
This indicates a prevalence of middleboxes that mirror MPTCP options sent in SYN in our \zmap scans.
On port 443, the phenomenon is much more prominent in IPv4, where almost 80\% of the sender's keys are mirrored compared to $\approx$8\% of keys on IPv6.
On port 80 (not shown), we find that middlebox interference is even more elevated, with almost 90\% and 30\% of received sender's keys identified as being mirrored for IPv4 and IPv6, respectively.
Note that MPTCPv1 hosts in our scans are unlikely to be affected by mirroring middleboxes since the sender's key is no longer sent in the SYN  for this protocol version (refer to \cref{sec:zmapMethod}).
On the other hand, for MPTCPv0, since we receive a legitimate sender's key along with an \texttt{MP\_CAPABLE} option in the SYN-ACK response (as per protocol's expectation), such responses can result in false-positives in our resulting analysis.
Furthermore, for our MPTCPv1 scans, we filter out responses that simply replay the MPTCP options sent in the SYN and do not explicitly report support for version 1 in SYN-ACK.

\begin{figure}[t!]
	\centering
	\includegraphics[width=\columnwidth]{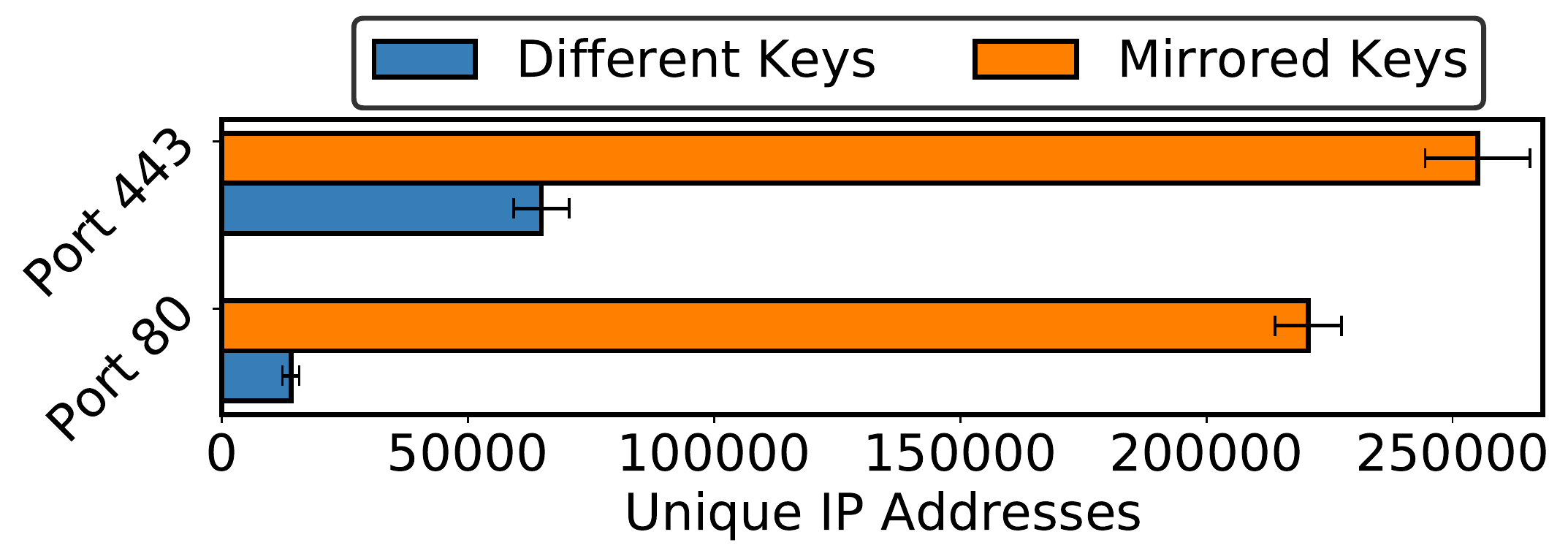}
	\caption{Unique IPv4 addresses that returned MPTCPv0 options in our \zmap scans over port 80 and 443 categorized by the sender's key.}
	\label{fig:ipv4zmap-key}
\end{figure}

We now analyze the share of hosts that are affected by middleboxes mirroring MPTCPv0 options in our measurements.
\Cref{fig:ipv4zmap-key} shows the aggregate number of unique potential MPTCP targets scanned over IPv4 for which the sender's key was mirrored (in orange) and different from ours (in blue) for ports 80 and 443.
It is evident that middleboxes affect \zmap scans significantly as many hosts on both ports have mirrored keys.
Interestingly, we find that the presence of middleboxes is far greater on port 80 than on port 443, as port 80 has 96\% of hosts with mirrored keys compared to 81\% on port 443.
In contrast, 6484 and 42294 hosts send back different sender's keys on ports 80 and 443, respectively.
For IPv6, we received different sender's key responses from 31 IP addresses on port 80 and 157 on port 443 (not shown).

The result is quite intriguing as it hints at HTTPS having far more support for MPTCPv0 than HTTP over both IPv4 and IPv6.
We also investigate whether any hosts that are middlebox-affected on port 80 are MPTCP-capable on port 443, but we find no such intersection.
This leads us to believe the following contrasting possibilities.
First, HTTPS traffic is end-to-end encrypted at the application layer; it is possible that a large number of middleboxes do not modify the transport layer options of user traffic.
This results in a much smaller percentage of hosts that are affected by middleboxes that inject replayed TCP extensions.
Second, the result may still include non-MPTCP end-hosts, which are affected by middleboxes that also modify the sender's key value of SYN-ACK packets.

\begin{tcolorbox}[title=\textit{Takeaway}, enhanced, breakable]
	Our \zmap scans reveal significant support for MPTCP over IPv4 compared to IPv6.
	MPTCPv1 support on both IPv4 and IPv6 was almost non-existent before October 2021 and remains to be minimal compared to MPTCPv0.
	However, \zmap includes a large share of seemingly MPTCP-capable addresses, where middleboxes are mirroring TCP options, which significantly impacts the accuracy of MPTCPv0 scans.
	For IP addresses that responded with different options, we find potential MPTCP support to lean more towards HTTPS than HTTP for both IPv4 and IPv6.
\end{tcolorbox}

\subsection{Finding Interfering Middleboxes}\label{sec:tracebox}

While our ZMap analysis in the previous section filters out hosts that are affected by middleboxes simply mirroring TCP options, it still does not entirely capture the true state of MPTCP deployment in the Internet.
First, our filtering mechanism assumes that all hosts which reply with the same sender's key as ours are middlebox-affected and \emph{do not} support MPTCP.
However, this excludes the possibility of legitimate MPTCP hosts whose MPTCP options in the SYN-ACK are either stripped or overwritten by middleboxes -- thus resulting in false negatives.
Second, our analysis may also include false positives due to middleboxes that may perform complex operations on packets with extended TCP options, \eg modifying sender's keys.
Note that the assumption does not affect MPTCPv1 \zmap scans as the sender's key is excluded in the MPTCPv1 SYN and, therefore, cannot be modified by middleboxes.
Therefore, we detect the presence of interfering middleboxes by running \tracebox~\cite{tracebox} towards targets that sent the \texttt{MP\_CAPABLE} option in our MPTCPv0 \zmap scans.%

\smallskip
\noindent \textbf{Methodology.}
Similar to our \zmap methodology, we issue \tracebox requests with the \texttt{MP\_CAPABLE} option towards a target address.
In the reply, we receive responses from intermediate routers on the path, including any modifications made.
Overall, we observe the following different behaviors in \tracebox responses.

\begin{enumerate}[label=\Roman*., nolistsep,labelwidth=!, labelindent=0pt]
	\item Only the target IP modifies the \texttt{MP\_CAPABLE} option.
	\item An intermediate hop modifies the \texttt{MP\_CAPABLE} option.
	\item The target was unresponsive or the query timed out.
\end{enumerate}

Based on these three categories, we classify MPTCP support as follows.
Since
\emph{category I} responses are caused by IP addresses updating MPTCP options with their own sender's key in the SYN-ACK response; we classify such targets
as \emph{truly} MPTCP-capable.
Targets in \emph{category II}
are clearly affected by middleboxes on the path and hence tagged as
\emph{middlebox-affected}.
Lastly, we classify hosts in \emph{category III} as \emph{unreachable}.

\smallskip
\noindent \textbf{True MPTCP support in the Internet.}
We check if any of the end-hosts that mirror our MPTCP key in \zmap (mirrored key hosts in \Cref{fig:ipv4zmap-key}) truly support MPTCP.
Interestingly, we did not \emph{any} end-host among these that sent back the \texttt{MP\_CAPABLE} flag (indicating MPTCP support) to our \tracebox probe for both IPv4 and IPv6.
This confirms that our \zmap analysis does not lead to any false negatives, and checking for mirrored MPTCP options in SYN-ACK is an effective first step in filtering out middleboxes.

\begin{figure}[t!]
	\centering
	\includegraphics[width=\columnwidth]{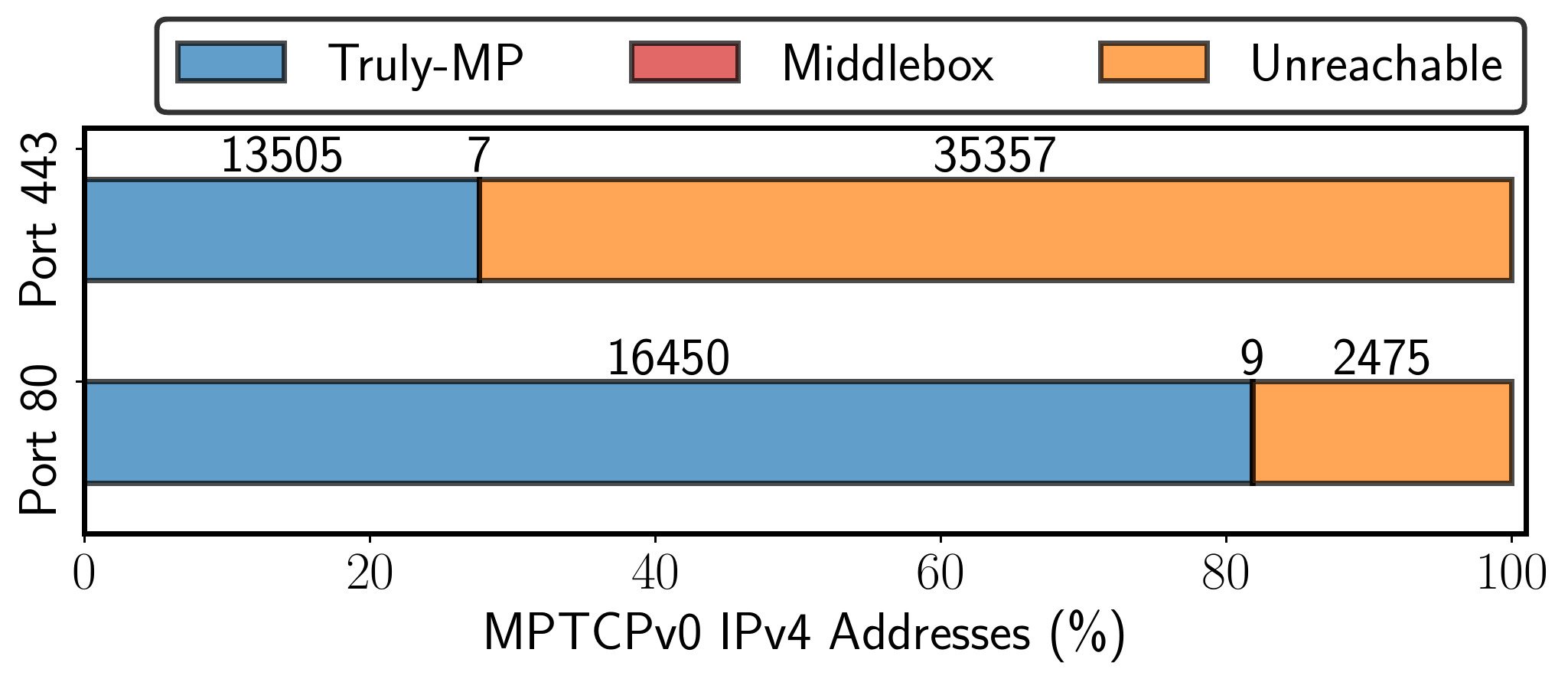}
	\caption{\tracebox analysis for \emph{consistently} responsive IPv4 addresses over port 80/443. The blue region denotes IP addresses that \emph{truly} support MPTCP, red (barely visible) are IP addresses affected by middleboxes on the path, and orange are unreachable.}
	\label{fig:ipv4tracebox}
\end{figure}

We continue our analysis with \tracebox responses from hosts that respond with different sender's keys in \zmap. %
We observe that a large percentage of targets do not respond to our \tracebox queries and are therefore classified as \emph{unreachable}.
This behavior is slightly more predominant in IPv4 than in IPv6, primarily due to the different target sets (Internet-wide in IPv4, hitlist-based in IPv6).
In IPv6, we only see unreachable hosts on port 443 ($\approx$ 82\%), where the majority of targets are located in the same prefix of a Dutch ISP.
These IPv6 targets respond to our \tracebox queries with \textit{Destination Unreachable (administratively prohibited)}, which hints at blocking of our queries by the ISP.
In IPv4, the number of \emph{unreachable} targets is significantly higher on port 443 ($\approx$ 90\%) than on port 80 ($\approx$~48\%).
As a result,
we only target addresses that were \emph{consistently reachable}, responded with \texttt{MP\_CAPABLE} and sent a different MPTCP sender's key in SYN-ACK at least once in the last three months of our measurement period, i.e., from October to December 2021.
This precaution deflates the number of \emph{unreachable} hosts from further analysis as it removes transient hosts that are only active for short periods of time.
In total, we ran \tracebox towards $\approx$ 20k addresses for port 80 and 48k addresses for port 443 in IPv4.
In IPv6, we sent \tracebox queries to 1267 and 1585 addresses for TCP/80 and TCP/443, respectively.
We confirm that the number of \emph{truly} MPTCP-capable addresses remains comparable before and after pruning transient IPs.
\Cref{fig:ipv4tracebox} shows the overview of our \tracebox analysis for IPv4.

Despite reducing the input dataset, we observe that a large share of IPv4 targets over port 443 are still \emph{unreachable} ($\approx$ 72\%) and the absolute number of responsive addresses is similar on both port 80 and 443.
Contrary to our initial assessment based on the \zmap results, we find \emph{true} MPTCP support in IPv4 to be slightly higher on port 80 (16.5k hosts) than on port 443 (13.5k hosts).
We also find that the \emph{true} support has almost doubled within the year 2021 in IPv4, as our previous observations found only 7.5k hosts and 6.9k hosts to \emph{truly} support MPTCP over port 80 and 443, respectively, at the end of December 2020~\cite{aschenbrennersingle}.
Similarly, in IPv6, we find that 1195 and 1184 hosts \emph{truly} support MPTCP over TCP/80 and TCP/443, respectively.
This is an increase of almost 40$\times$ in MPTCP support compared to December 2020 (31 on TCP/80 and 27 on TCP/443).

Our results also show a limited interference of middleboxes affecting MPTCP support in IPv4.
In IPv4, we find 9 and 7 hosts that are affected by middleboxes on the path on TCP/80 and TCP/443, respectively.
For all recorded instances, the middleboxes stripped the MPTCP options from the SYN packet.
Note that the result is in contrast to our previous observations~\cite{aschenbrennersingle} in December 2020 where we found 402 and 1.3k \emph{middlebox-affected} end-hosts on port 80 and 443, including 6 \emph{truly} MPTCP hosts. 
While we are unsure of the exact reason behind the significant decrease in hosts affected by middleboxes, we find that none of the \emph{middlebox-affected} end-hosts in our previous analysis responded to our recent scans in late 2021. 
As a result, we likely attribute those short-lived end-hosts to enthusiasts, system tinkerers, or researchers that may have used MPTCP for short time periods. 
Note that since MPTCP options are stripped for the \emph{middlebox-affected} end-hosts, we cannot accurately assess the true support for MPTCP within this group.
On the other hand, similar to our past observations, we do not find a single middlebox-affected target address in IPv6.

We attempt to investigate the deployment nature of middleboxes that impact MPTCP traffic using Nmap~\cite{nmap} fingerprinting.
Unfortunately, this did not lead to fruitful results due to the following hindrances.
First, the majority of middleboxes that impact our study do not respond to \tracebox probes, and hence we are unable to identify their IP address.
Second, for the handful of middleboxes that we positively identified, the accuracy of Nmap is too low to confidently identify their hardware and OS characteristics.
However, we find that most middleboxes we identified in our latest scans in 2021 belonged to three different ISPs operating in the USA, UK and Asia.
We leave the thorough analysis of middleboxes in the Internet and their impact on TCP extensions to future work.

\begin{figure}[t!]
	\centering
	\includegraphics[width=\columnwidth]{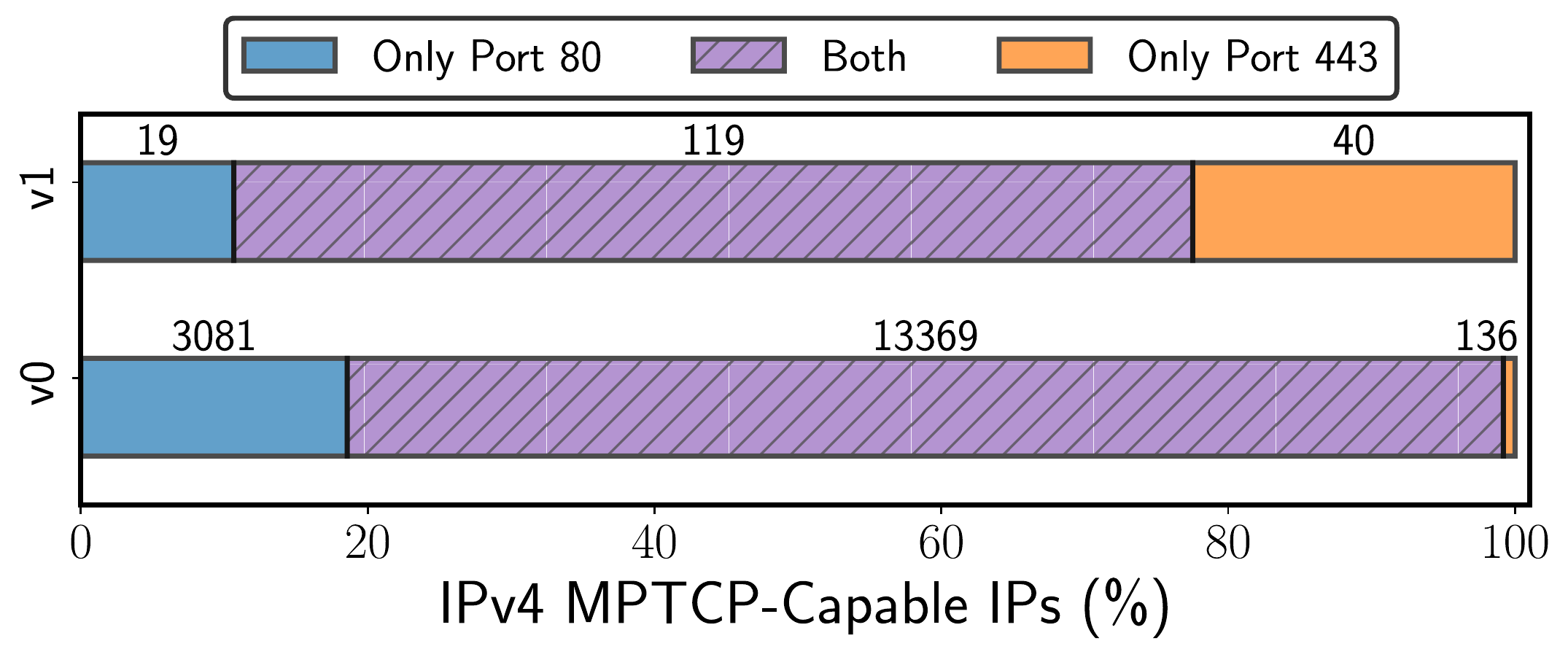}
	\caption{MPTCP (v0 and v1) support for HTTP and HTTPS in IPv4.}
	\label{fig:ipv4-port-breakup}
\end{figure}

\smallskip
\noindent \textbf{MPTCP Port Overlap.}
Since an MPTCP-capable machine can offer different services concurrently, we now examine the overlap between TCP/80 and TCP/443 end-hosts.
\Cref{fig:ipv4-port-breakup} shows the port breakup of all MPTCPv0 and MPTCPv1 capable IPv4 addresses throughout our study over either port 80/443.
As shown in the figure, most IPv4 MPTCP hosts provide services over both ports simultaneously.
It must be noted that this result is in stark contrast to our previous observations made in December 2020, in which we found that the majority of IPv4 addresses provided complementary services over either of the two ports.
At the end of December 2021, we find that 80.6\% and 66.9\% of IPv4 addresses support both HTTP and HTTPS simultaneously over MPTCPv0 and MPTCPv1, respectively.
Hosts that only support HTTPS are in the minority in MPTCPv0 ($<$ 1\%) as more addresses tend to provide services over HTTP (18.5\%).
On the other hand, HTTPS-only support is more popular over MPTCPv1 (22.4\%) compared to HTTP-only support (10.7\%).
The picture is different for IPv6, where more than 77\% of addresses support MPTCPv0 on both ports and 100\% of addresses support MPTCPv1 on both ports.

\begin{figure}[t!]
	\centering
	\includegraphics[width=\columnwidth]{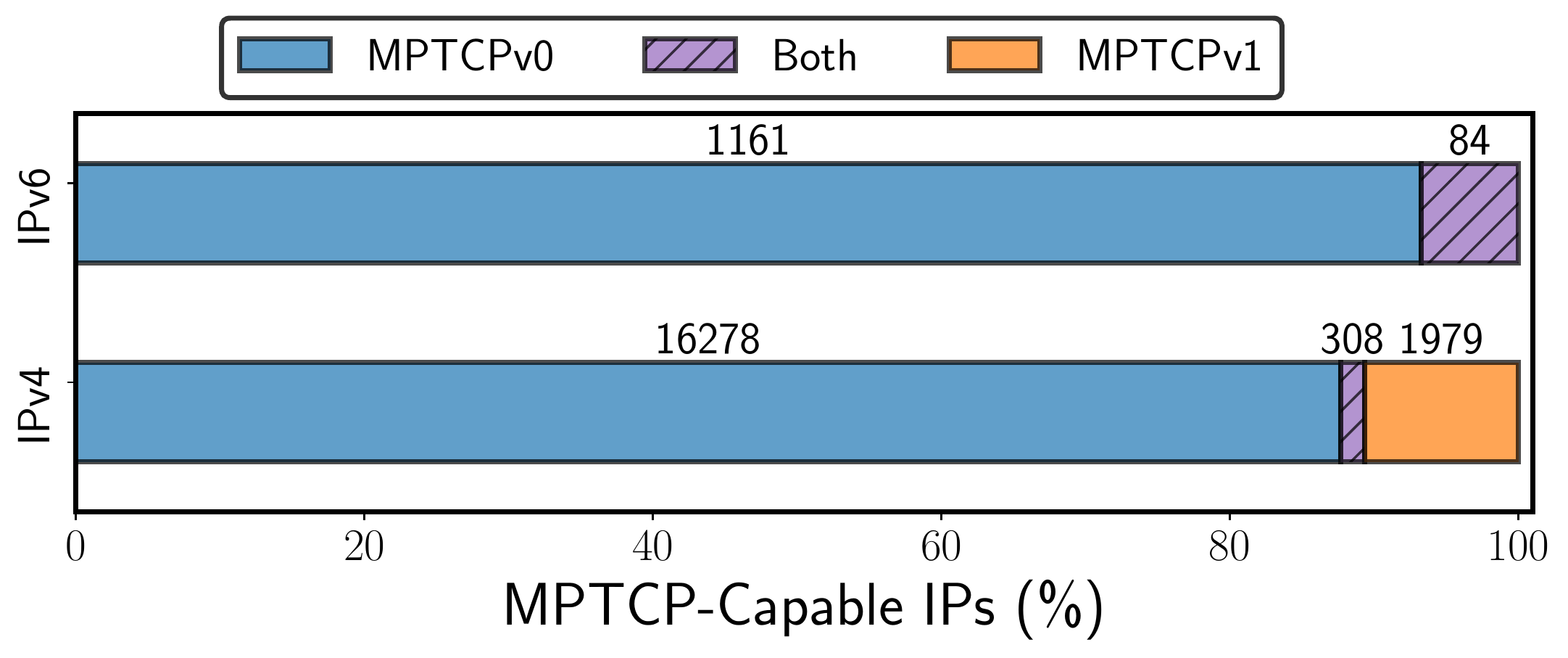}
	\caption{Overlap between MPTCPv0 and MPTCPv1 support in IPv4 and IPv6.}
	\label{fig:ipv4-version-breakup}
\end{figure}

\smallskip
\noindent \textbf{MPTCP Version Overlap.}
Considering the increasing adoption trend of MPTCPv0 and MPTCPv1 within 2021 (as shown in \cref{sec:mptcp-zmap}), we now take a deeper look into understanding the deployment nature of both versions.
\Cref{fig:ipv4-version-breakup} shows the overlap between end-hosts that support MPTCPv0 and MPTCPv1. 
For this analysis, we combine end-targets that were active on either TCP/80 or TCP/443 in our measurements.
We observe that a large majority of the end-hosts on IPv4 and IPv6, i.e., 87.68\% and 93.25\% respectively, only support the earlier experimental MPTCPv0 protocol.
Despite being available in the default Linux kernel, just 10.65\% of IPv4 addresses provide services only on MPTCPv1.
On the other hand, all IPv6 end-hosts that support MPTCPv1 also support MPTCPv0---with all of these addresses belonging to Apple (AS6185).
Similarly, we find that a fraction of IPv4 addresses support both MPTCPv0 and MPTCPv1 simultaneously (1.66\%).
Further investigation reveals that the majority of these overlapping MPTCP targets in IPv4 belong to Apple (AS6185) and Telecom Italia (AS3269). 

\begin{figure}[t!]
	\centering
	\begin{subfigure}{0.55\columnwidth}
		\includegraphics[width=\columnwidth]{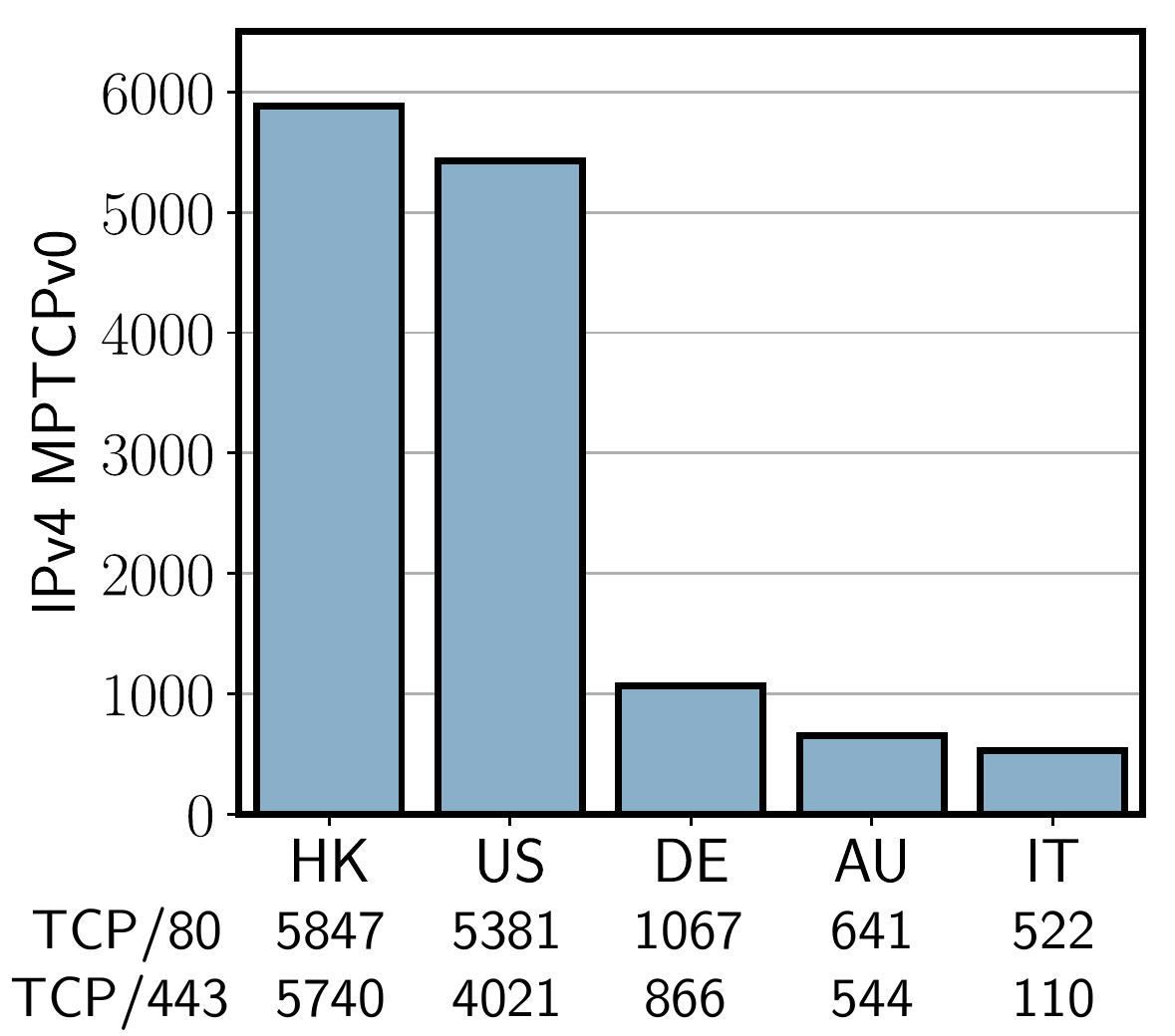}
		\caption{MPTCPv0}
		\label{fig:ipv4tracebox-country}
	\end{subfigure}%
	\begin{subfigure}{0.435\columnwidth}
		\includegraphics[width=\columnwidth]{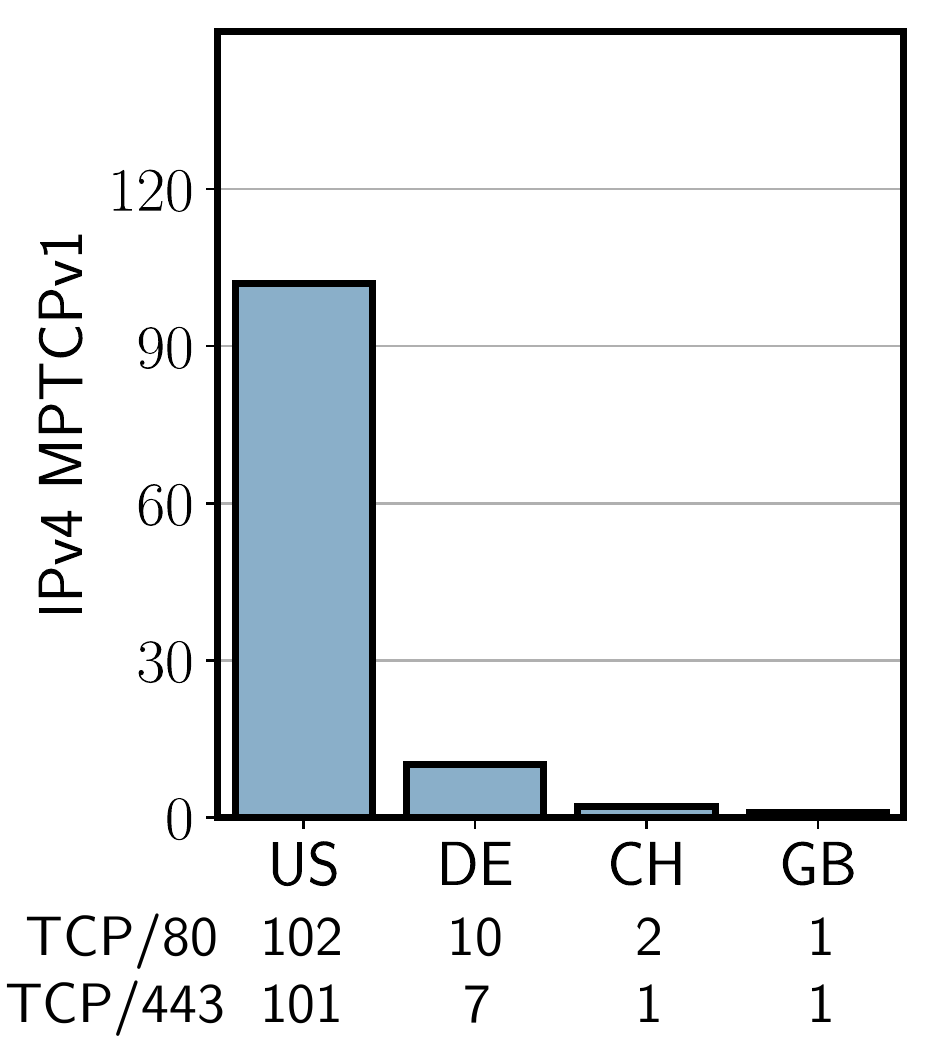}
		\caption{MPTCPv1}
		\label{fig:ipv4-mptcpv1-country}
	\end{subfigure}
	\caption{Geographic distribution of \emph{truly} MPTCP-capable IPv4 addresses verified by \tracebox. The bars show counts of unique IPs over both port 80 and 443 (including common IPs). The numbers below the x-axis denote the MPTCP-capable IPs serving over port 80 and 443 in that country.}
	\label{fig:ipv4-geography}
\end{figure}

\begin{tcolorbox}[title=\textit{Takeaway}, enhanced, breakable]
	Using \tracebox we find that the \emph{true} support for MPTCP is much lower compared to numbers reported by \zmap, around 16.5k and 13.5k for IPv4 and 1195 and 1184 for IPv6 over HTTP and HTTPS, respectively. Compared to our last observations in December 2020, MPTCP support has grown significantly ($\approx$2$\times$ in IPv4 and $\approx$40$\times$ in IPv6) in the year 2021. Despite its experimental nature, MPTCPv0 is still the more popularly deployed version on IPv4 and IPv6 and a majority of MPTCP hosts (both v0 and v1) provide services over port 80 and 443 simultaneously. We also find a few MPTCP end-hosts to be middlebox-affected over IPv4, while IPv6 MPTCP support remains largely unaffected. 
\end{tcolorbox}

\subsection{Geo-distribution of MPTCP-capable hosts} \label{sec:geo-distribution}

We now shed some light on the physical deployment locations and operational zones of end-hosts that \emph{truly} support MPTCP.
\Cref{fig:ipv4tracebox-country} visualizes the top 5 countries with the most MPTCPv0 capable host densities while \Cref{fig:ipv4-mptcpv1-country} shows the top-4 countries with the most MPTCPv1 capable hosts on IPv4, arranged in decreasing fashion.
We use the MaxMind database~\cite{maxmind} for our analysis
and only show country-level breakup since more fine-granular IP geo-location suffers from accuracy shortcomings \cite{livadariu2020accuracy,scheitle2017hloc,shavitt2011geolocation}.
\Cref{tbl:asn-ipv4-tracebox} provides further insights, as it shows the top 10 ASes with most IPv4 MPTCPv0 hosts on port 80 and 443.
Similarly, \Cref{tbl:asn-ipv4-v1} shows the top-5 ASes with MPTCPv1 hosts on both ports in IPv4.
The tables also list the associated organization name, country and AS rank, which we obtain from CAIDA's AS database~\cite{as_ranks_site}.

\begin{table}[!t]
	\setlength{\extrarowheight}{0pt}
	\addtolength{\extrarowheight}{\aboverulesep}
	\setlength{\aboverulesep}{0pt}
	\setlength{\belowrulesep}{0pt}
	\begin{adjustbox}{width=\columnwidth}
		\centering
		\begin{tabularx}{\columnwidth}{llllll}
			\toprule
			\textbf{ASN}~                           & \textbf{\#Port80} & \textbf{\#Port443} & \textbf{Rank} & \textbf{Country} & \textbf{Owner} \\
			\midrule
			9269                                    & 6461              & 6370               & 364           & HK               & HK Broadband   \\
			\rowcolor[rgb]{0.753,0.753,0.753} 6185  & 1347                & 1344               & 13577         & US               & Apple Inc.     \\
			61157                                   & 674               & 534                & 1368          & DE               & Plus Server    \\
			\rowcolor[rgb]{0.753,0.753,0.753} 1221  & 529               & 456                & 76            & AU               & Telstra Corp.  \\
			18618                                   & 390               & 390                & 3915          & US               & West Central   \\
			\rowcolor[rgb]{0.753,0.753,0.753} 18943 & 353               & 352                & 3533          & US               & Yelcot Tele.   \\
			7922                                    & 419               & 209                & 27            & US               & Comcast        \\
			\rowcolor[rgb]{0.753,0.753,0.753} 11976 & 232               & 231                & 2360          & US               & Fidelty Comm.  \\
			202870                                  & 404               & 2                & 16712         & IT               & Dimensione     \\
			\rowcolor[rgb]{0.753,0.753,0.753} 15129 & 369               & 1                  & 4034          & US               & Geneseo Tele.  \\
			\bottomrule
		\end{tabularx}
	\end{adjustbox}
	\caption{Top 10 Autonomous Systems for \emph{truly} MPTCP-enabled hosts in IPv4.}
	\label{tbl:asn-ipv4-tracebox}
\end{table}

\begin{table}[!t]
	\setlength{\extrarowheight}{0pt}
	\addtolength{\extrarowheight}{\aboverulesep}
	\setlength{\aboverulesep}{0pt}
	\setlength{\belowrulesep}{0pt}
	\begin{adjustbox}{width=\columnwidth}
		\centering
		\begin{tabularx}{\columnwidth}{llllll}
			\toprule
			\textbf{ASN}~                            & \textbf{\#Port80} & \textbf{\#Port443} & \textbf{Rank} & \textbf{Country} & \textbf{Owner} \\
			\midrule
			6185                                     & 98                & 98                 & 14234         & US               & Apple Inc.     \\
			\rowcolor[rgb]{0.753,0.753,0.753} 396986 & 2                 & 2                  & 16570         & US               & Bytedance      \\
			206293                                   & 0                 & 3                  & 22645         & DE               & ProIO GmbH ~~~ \\
			\rowcolor[rgb]{0.753,0.753,0.753} 714    & 1                 & 2                  & 6949          & US               & Apple Inc.     \\
			3209                                     & 0                 & 2                  & 221           & DE               & Vodafone       \\
			\bottomrule
		\end{tabularx}
	\end{adjustbox}
	\caption{Top 5 Autonomous Systems for MPTCPv1 hosts in IPv4.}
	\label{tbl:asn-ipv4-v1}
\end{table}

\begin{table}[!t]
	\setlength{\extrarowheight}{0pt}
	\addtolength{\extrarowheight}{\aboverulesep}
	\setlength{\aboverulesep}{0pt}
	\setlength{\belowrulesep}{0pt}
	\begin{adjustbox}{width=\columnwidth}
		\centering
		\begin{tabularx}{\columnwidth}{llllll}
			\toprule
			\textbf{ASN}~                           & \textbf{\#Port80} & \textbf{\#Port443} & \textbf{Rank} & \textbf{Country} & \textbf{Owner} \\
			\midrule
			6185                                    & 1163              & 1163               & 14234         & US               & Apple Inc.     \\
			\rowcolor[rgb]{0.753,0.753,0.753} 63949 & 4                 & 3                  & 7042          & US               & Linode         \\
			4811                                    & 3                 & 3                  & 2034          & CN               & China Tele.    \\
			\rowcolor[rgb]{0.753,0.753,0.753}   714 & 2                 & 2                  & 6949          & US               & Apple Inc.     \\
			201155                                  & 2                 & 2                  & 26184         & CH               & EmbeDD         \\
			\bottomrule
		\end{tabularx}
	\end{adjustbox}
	\caption{Top 5 Autonomous Systems for \emph{truly} MPTCPv0-enabled hosts in IPv6.}
	\label{tbl:asn-ipv6-tracebox}
\end{table}

We find that close to half of all IPv4 MPTCP hosts are using MPTCPv0 and are deployed in Hong Kong, totaling more than 6000 unique addresses over port 80 and port 443.
Comparing our results to our observations drawn in December 2021~\cite{aschenbrennersingle}, we find that the majority of MPTCP hosts in Hong Kong started operation in the year 2021 and are hosted by a major ISP in the region, Hong Kong Broadband (AS9269). 
However, HK Broadband has almost negligible MPTCPv1 support in IPv4, which is still being dominated by the USA.
Furthermore, the USA also comes close second as the largest supporter of either MPTCPv0 or MPTCPv1, with almost 5.8k unique MPTCP-capable hosts.
\Cref{tbl:asn-ipv4-tracebox} shows that within the US, Apple has the largest deployment of MPTCP servers operational on both port 80 and 443, totaling close to 3000 unique IPv4 addresses.
The result is unsurprising since Apple has been publicly known to use MPTCP for several iOS services, \eg Siri, Music, Maps, and has recently allowed third-party developers to utilize MPTCP for non-system-native apps~\cite{apple_backup}.
We also find that Apple has recently started supporting the newly standardized MPTCPv1, even migrating its existing MPTCPv0 hosts to MPTCPv1 (we investigate this further in \Cref{sec:discussion:sub:apple}).
The third-largest support for MPTCP over IPv4 comes from Germany, mainly due to servers hosted by Plus Server, a major cloud hosting company in the region.
We also observe that many network operators and ISPs across the globe (specifically in the US) are utilizing MPTCP within their networks to enhance several of their client-facing services.
Interestingly, Korea Telecom, which publicly announced exploiting MPTCP to provide Gigabit speeds over Wi-Fi and LTE~\cite{kt-gigalte} fails to be among the top-10 ASNs with MPTCP-capable hosts (with 205 unique MPTCP addresses).
Interestingly, we also observe from Figure~\ref{fig:ipv4tracebox-country} that in certain deployments such as Dimensione (AS202870) and Geneseo Telecom (AS15129), MPTCP deployment favors one port over the other, showcasing an organization's tendency to utilize MPTCP for serving specific application traffic.

In \Cref{tbl:asn-ipv6-tracebox} we show the AS distribution of \emph{truly} MPTCPv0-capable IPv6 addresses.
Compared to IPv4, MPTCPv0 support in IPv6 is completely dominated by Apple deployment with more than 1k addresses.
The rest of the top 5 ASes in IPv6 only has a handful of MPTCPv0 capable hosts.
We omit the top 5 table for IPv6 MPTCPv1, as it only contains a single entry, namely Apple with 84 IPs.
Overall, we find that the current MPTCP deployment in IPv4 and IPv6 spans more than 80 countries across the globe.

\begin{tcolorbox}[title=\textit{Takeaway}, enhanced, breakable]
	The largest support for MPTCP in IPv4 comes from Hong Kong Broadband, which significantly favors MPTCPv0 over MPTCPv1. The second largest MPTCP deployment in IPv4, and largest in IPv6, is within the USA---backed by Apple and other ISPs in the country that use MPTCP to enhance their services.
\end{tcolorbox}
\subsection{Middlebox Impact on MPTCP Perceived Quality} \label{subsec:mptcpqoe}

Our analysis in \Cref{sec:tracebox} revealed a prevalence of middleboxes that modify extensions MPTCP relies on.  %
Previous research has shown that certain middleboxes, such as firewalls or load balancers, manipulate packets that do not fit pre-defined rule sets, \eg by marking them low-priority or forwarding them on longer paths~\cite{hesmans2013tcp}.
In this section, we want to answer whether middleboxes treat MPTCP application traffic differently from regular TCP traffic. 

We investigate this by initiating HTTP(S) GET requests using MPTCP %
from AWS in Germany towards IPv4 addresses that are marked \textit{potential}-MPTCP  in \Cref{sec:mptcp-zmap}.
We conduct the same measurements over regular TCP from the same data center in parallel.
For each successful GET response, we record \one the TCP handshake time (a.k.a. connect time), \two the TLS handshake time, \three time to first byte (TTFB), and \four the total completion time (roughly equates to website load time).
We run each measurement set, composed of 10+ runs, for almost two weeks. 
Overall, $\approx$~80\% and $\approx$~27\% targets responded to our GET requests on port 80 and 443, respectively.

 \begin{figure}[t]
	\begin{subfigure}{0.5\columnwidth}
    \includegraphics[width=\textwidth]{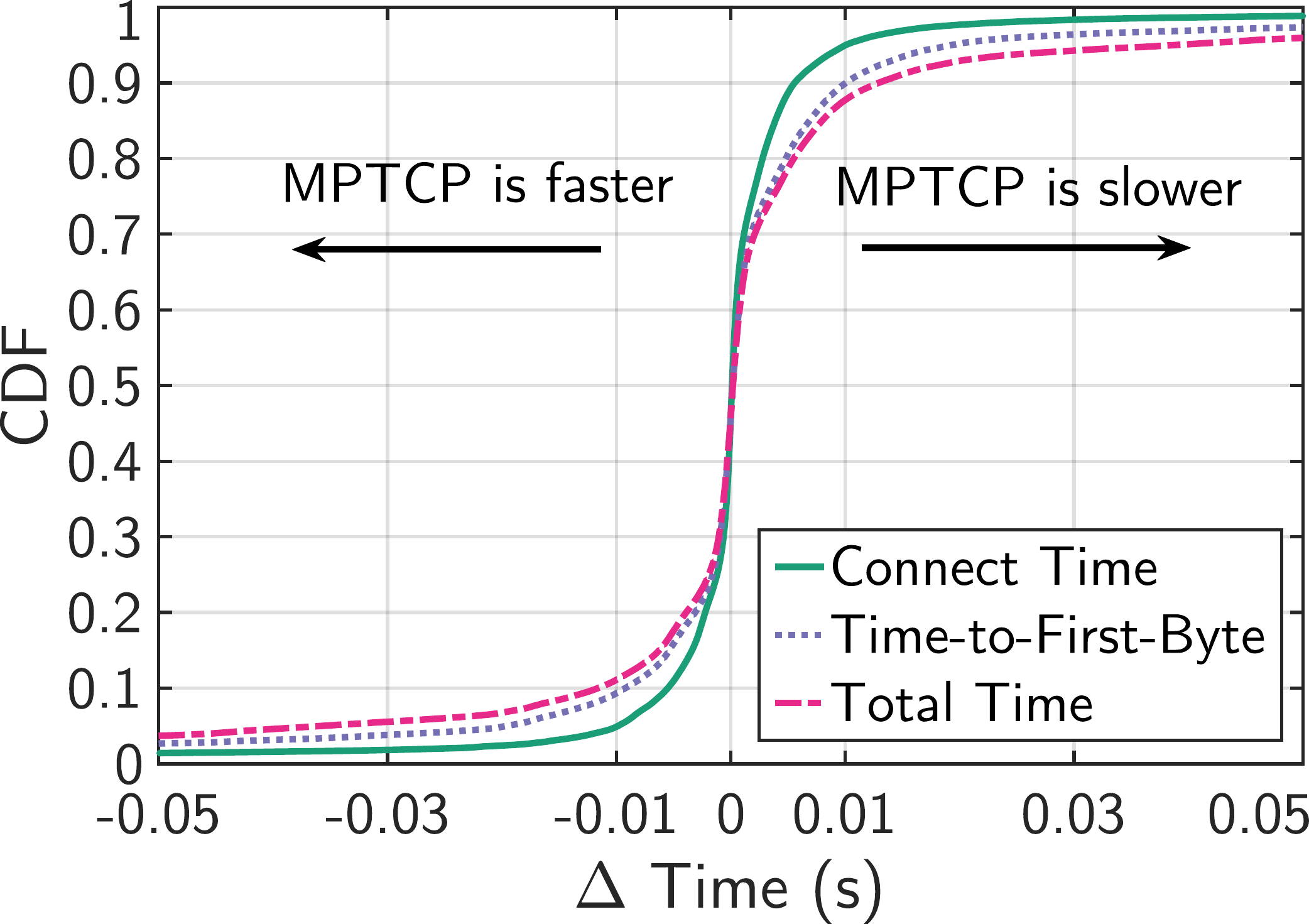}
\caption{}
    \label{fig:p80curl}
    \end{subfigure}%
    \begin{subfigure}{0.5\columnwidth}
    \includegraphics[width=\textwidth]{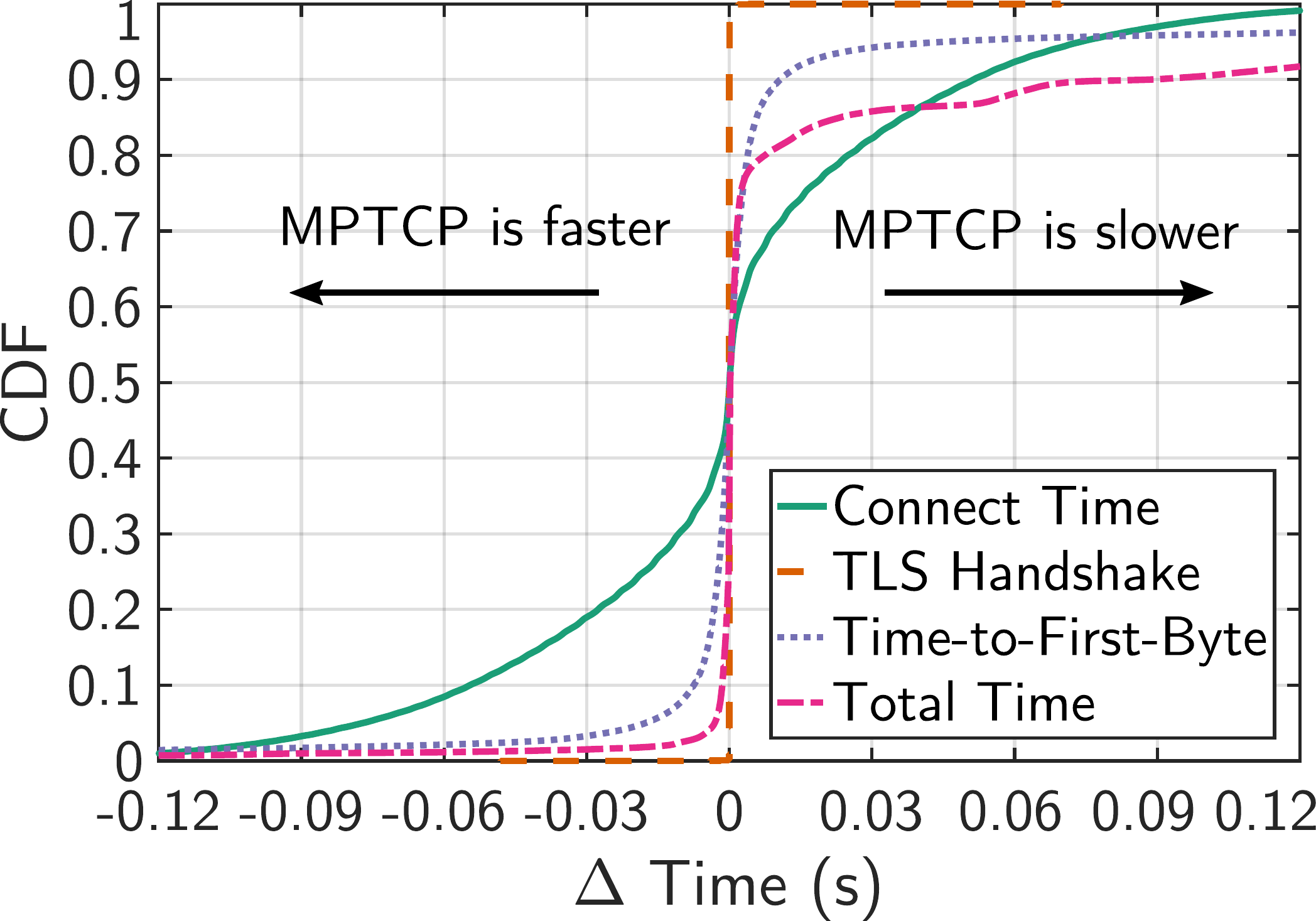}
\caption{}
    \label{fig:p443curl}
    \end{subfigure}
    \begin{subfigure}{0.5\columnwidth}
    \includegraphics[width=\textwidth]{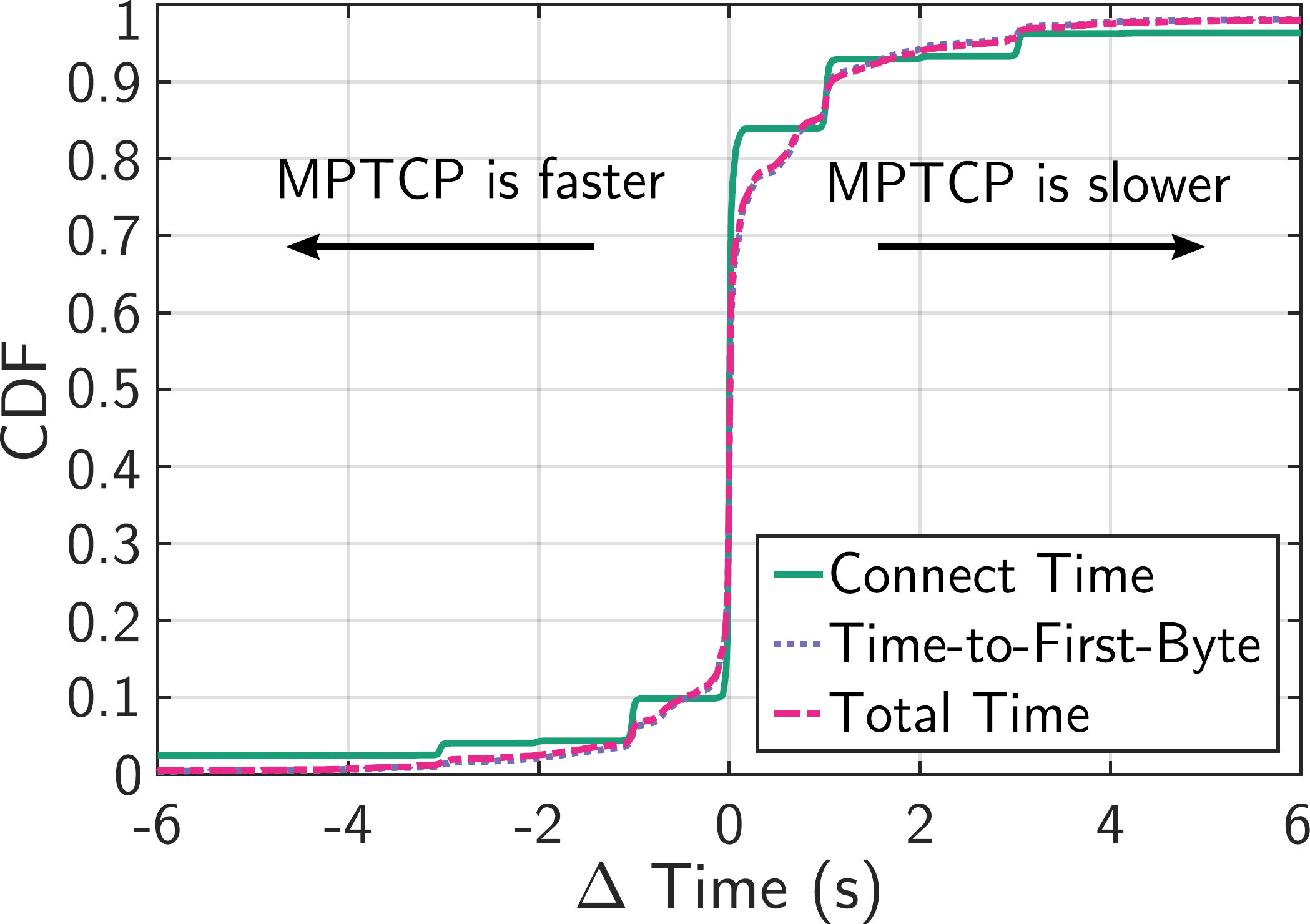}
\caption{}
    \label{fig:middleboxcurl}
    \end{subfigure}%
    	\begin{subfigure}{0.5\columnwidth}
    \includegraphics[width=\textwidth]{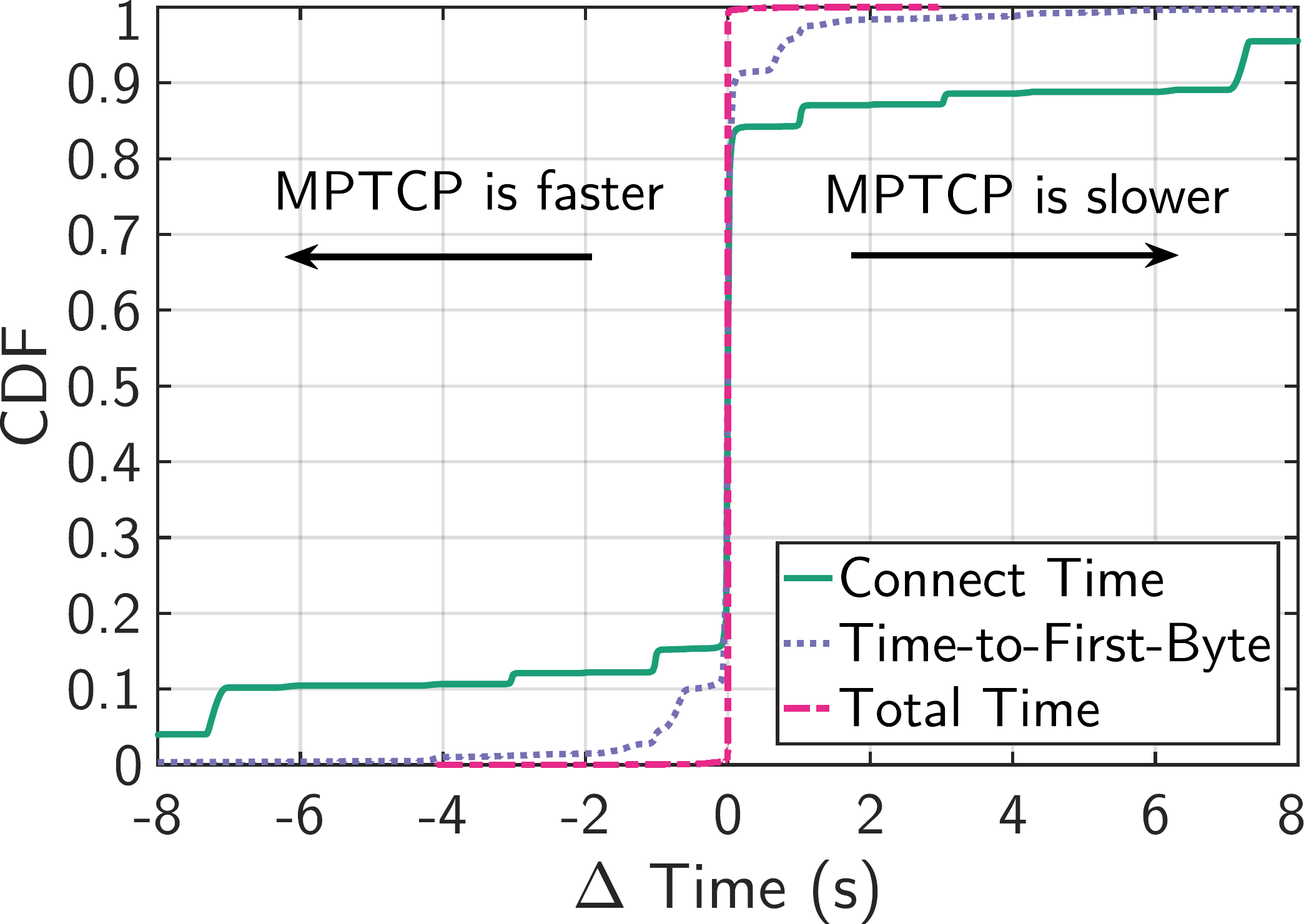}
\caption{}
    \label{fig:p80mpdiff0}
    \end{subfigure}%
    \caption{HTTP(S) GET timing deltas over MPTCP and TCP towards: \emph{truly}-MPTCP targets on port 80 (a) and port 443 (b); MPTCP targets affected by middleboxes via \tracebox (c); and targets affected by mirroring middleboxes (d). Please note the different x-axis time scale.}
    \label{fig:ipv4curlmiddlebox}
\end{figure}

\Cref{fig:p80curl,fig:p443curl} show the distribution of $\Delta$ time difference between responses from \emph{truly} MPTCP IPs identified in \Cref{sec:tracebox}.
Keep in mind that these targets are \emph{not} affected by middleboxes on the path.
$\Delta$ values less than zero denote targets that are faster using MPTCP, while $\Delta > 0$ are hosts that are faster over TCP. 
Values centered around zero indicate that both protocols perform similarly.
The symmetric upper and lower distributions in \Cref{fig:p80curl} shows that the clients observe no discernible difference using (MP)TCP if connecting to targets that support MPTCP over port 80.
MPTCP-capable targets on port 443 (shown in \Cref{fig:p443curl}) show similar results for all timing values except completion time, for which the distribution tilts slightly in favor of TCP. 

We now investigate the impact of middleboxes on MPTCP traffic. 
In \Cref{fig:middleboxcurl} we show the responses from MPTCP-capable targets found to be affected by middleboxes.
As can be observed, middleboxes treat MPTCP application traffic differently.
For $\approx$30\% of \emph{all} timing values, MPTCP is slower than TCP while TCP is slower for only 10\% of measurements.
Notice the difference in x-axis ticks of \Cref{fig:middleboxcurl} and \Cref{fig:p80curl}; indicating that middleboxes can expand TTFB and load time of MPTCP connections by several seconds. 
Likely, the MPTCP client falls back to TCP before initiating a data transfer for these targets since middleboxes strip away MPTCP options from the header~\cite{rfc8684}. 
As a result, such middleboxes only affect the TCP handshake phase, which also justifies large connect time values recorded for these targets.
However, not all middleboxes have a deleterious impact on MPTCP traffic, as seen in \Cref{fig:p80mpdiff0}.
The result shows that 
middleboxes that simply replay unknown TCP extensions (mirroring middleboxes) have no discernible effect on MPTCP traffic.
Keep in mind that data transfers over these connections end up using TCP since none of the end-targets in this group were found to support MPTCP.

\begin{tcolorbox}[title=\textit{Takeaway}, enhanced, breakable]
  We observe no significant difference in HTTP(S) GET responses when using MPTCP over TCP from \emph{truly} MPTCP-capable servers. 
  However, we find that certain middleboxes can aggressively delay MPTCP connections, whereas TCP remains largely unaffected.
\end{tcolorbox}

\section{MPTCP Internet Traffic Share}
\label{sec:mptcpTraffic}

We quantify the real-world MPTCP traffic share by analyzing two traffic traces from geographically diverse vantage points:
\one four years of traffic (from 2015 to 2019) on a Tier 1 ISP backbone link in North America (CAIDA traces~\cite{caida}) and
\two eight years of traffic (from 2014 to 2022) captured at the uplink of a Japanese university network (MAWI traces~\cite{mawi})

\begin{figure}[t!]
	\centering
	\begin{subfigure}{.49\columnwidth}
		\centering
		\includegraphics[width=\columnwidth]{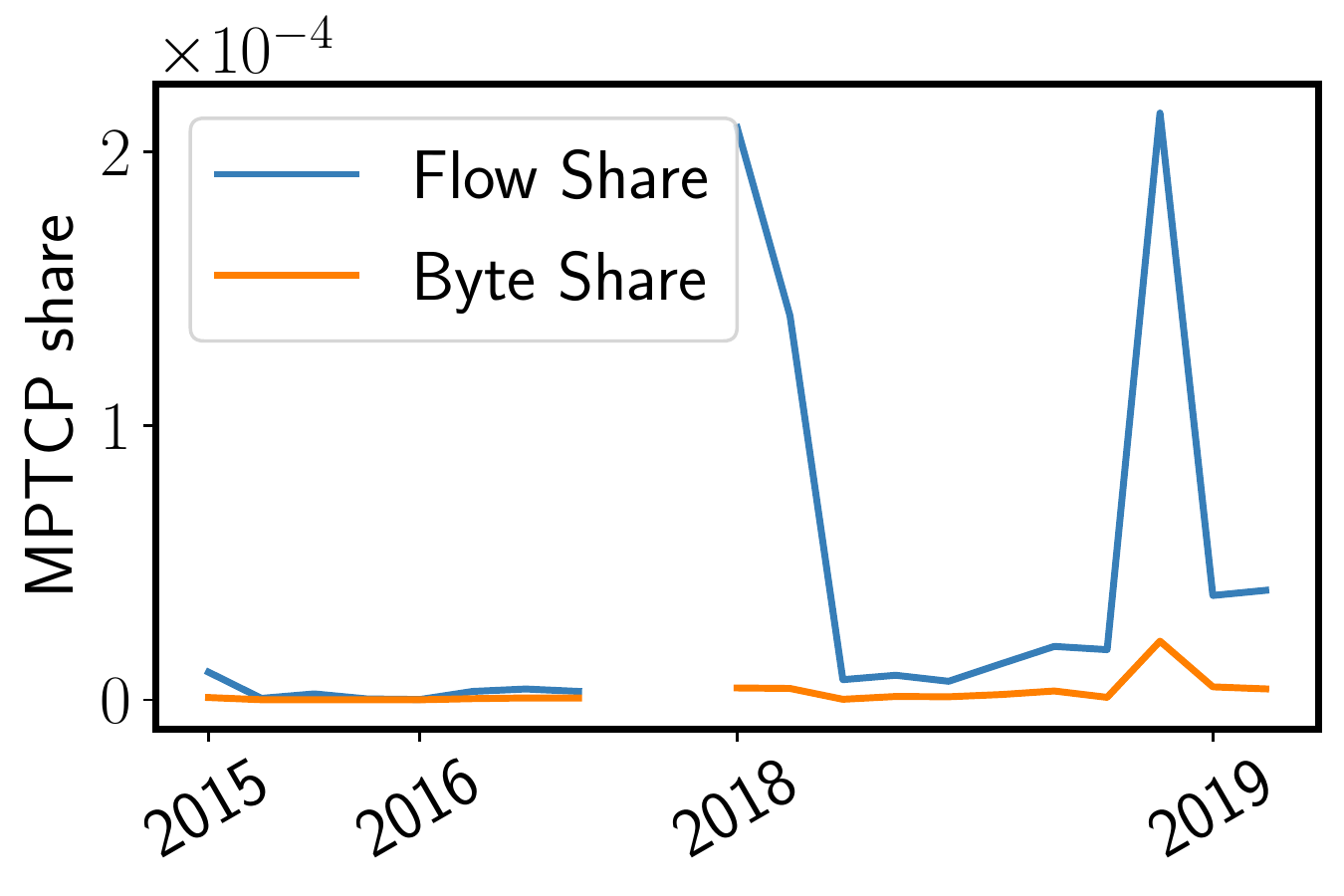}
		\caption{MPTCP traffic share.}
		\label{fig:caida-mptcp-share}
	\end{subfigure}%
	\begin{subfigure}{.51\columnwidth}
		\centering
		\includegraphics[width=\columnwidth]{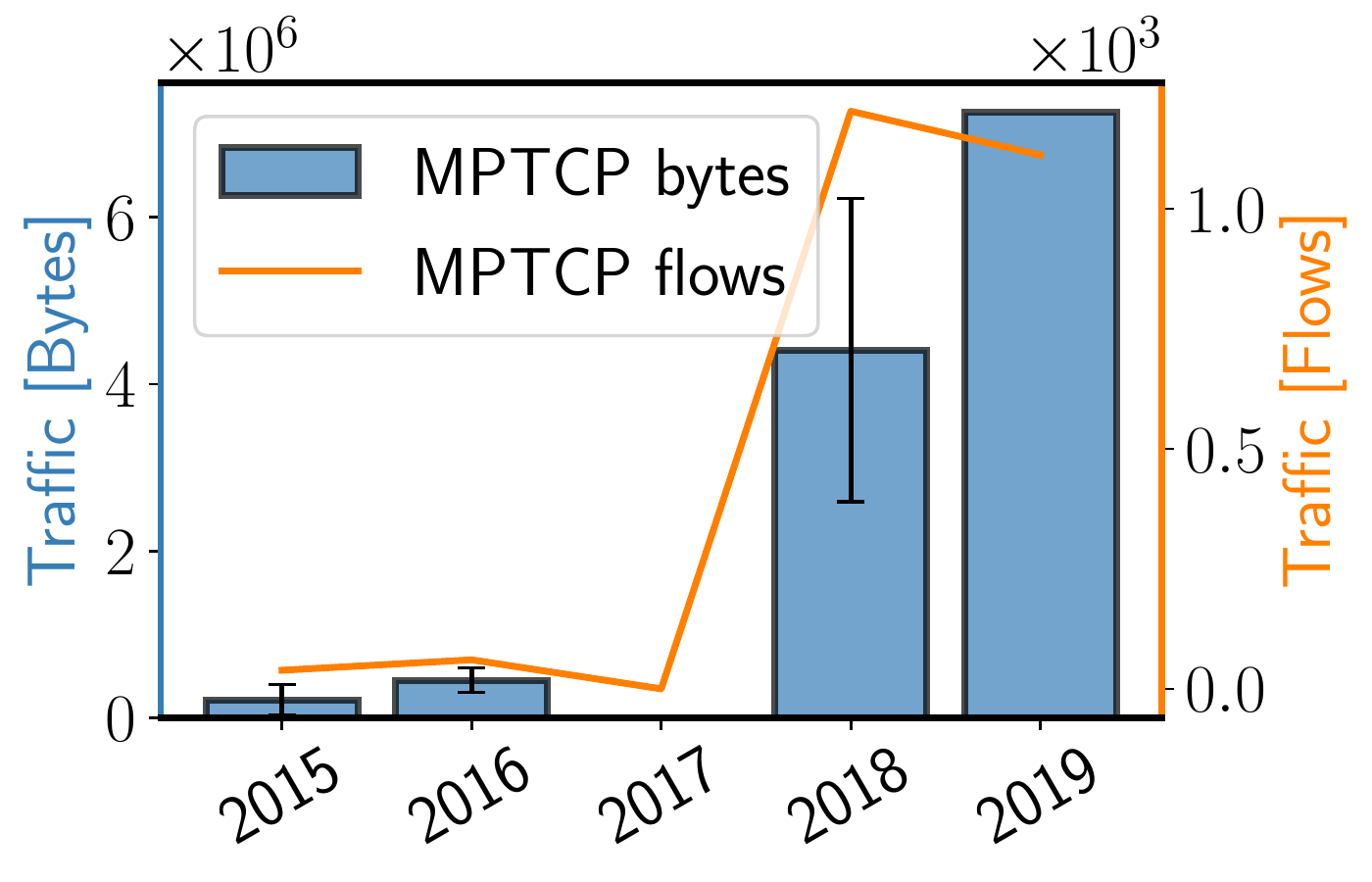}
		\caption{MPTCP absolute traffic.}
		\label{fig:caida-mptcp-absolute}
	\end{subfigure}%
	\caption{MPTCP traffic over time captured by CAIDA monitors on direction-A. (a) shows the share of MPTCP flows and bytes (compared to TCP) and (b) shows their absolute values across time. The gap is due to missing data for 2017.}
	\label{fig:caida-mptcp}
	\centering
\end{figure}

\noindent \textbf{CAIDA.} The CAIDA dataset includes bidirectional traffic captured at an Equinix data center connected to an ISP backbone link (we only consider single direction ``\textit{dir-A}" traffic in our analysis).
For 2015 and 2016, the monitor captures traffic of the ISP backbone connecting Chicago and Seattle, while for 2018 and 2019, the backbone links New York and São Paolo.
The dataset includes a one-hour trace per month for four months of 2015 and 2016 each, ten months for 2018 and January 2019.
No data is available for 2017 and after January 2019 since the monitored links have been upgraded to 100 Gbps and exceed capturing capacity.

\noindent \textbf{MAWI.}
The MAWI dataset includes traffic captured at samplepoint-F, a 1 Gbps transit link of the WIDE working group to an upstream ISP.
We analyze 15-minute captures of the third Thursday of each month, from January 2014 to December 2021.
This selection allows for better comparison between months, ensuring that weekday traffic is analyzed.

Generally, both CAIDA and MAWI datasets are anonymized, disallowing us to identify participating endpoints accurately.
This is sufficient to understand the popularity of MPTCP in real-world Internet traffic.
However, we analyzed a small set of unanonymized MAWI traces to better understand communication endpoints and compare them to our active measurements.
We remove all flows with less than five packet exchanges to prevent possible scanning traffic from influencing our study.

\subsection{MPTCP Traffic Characteristics} \label{sec:mptcpTrafficCharacteristics}

\begin{figure}[t!]
	\includegraphics[width=\columnwidth]{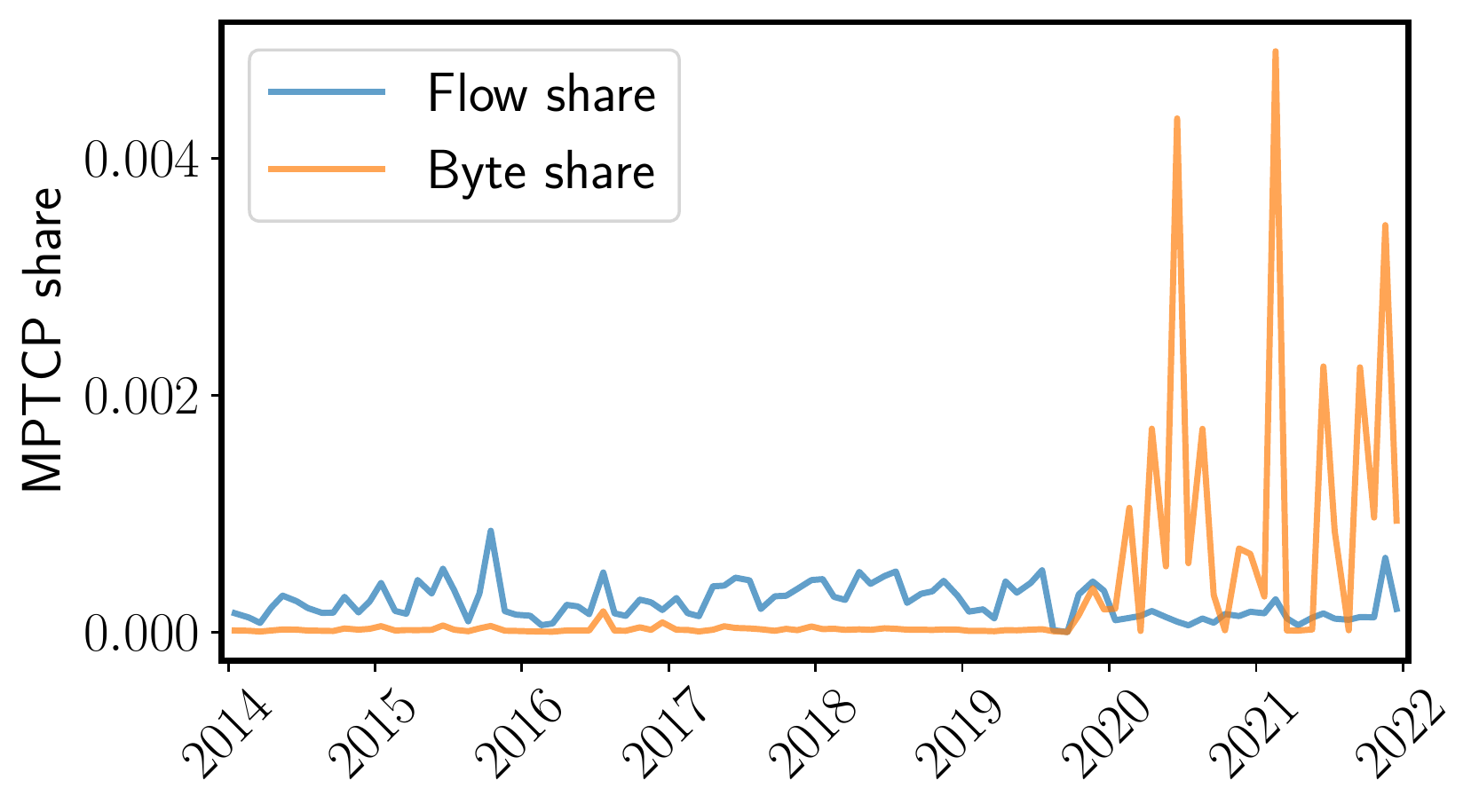}
	\caption{MPTCP traffic share for bytes and flows at MAWI's samplepoint-F.}
	\label{fig:mawi-mptcp-share}
\end{figure}

\Cref{fig:caida-mptcp-share,fig:mawi-mptcp-share} show the share of MPTCP flows and bytes over TCP at the CAIDA and MAWI vantage points, respectively.
We observe that the MPTCP share remains relatively consistently low in the CAIDA dataset, making up only 0.00006\% of TCP byte and 0.0003\% of TCP flow traffic.
However, there is an apparent uptick in MPTCP flow share at the start of 2018 that increases as the year progresses, reaching 0.005\%.
Interestingly, the trend is mostly missing on
MPTCP byte share, indicating a simultaneous rise of TCP traffic on the link.
By the end of 2018 (and the beginning of 2019), both MPTCP flow and byte share within the CAIDA dataset escalate significantly and peak at 0.02\% and 0.002\%, respectively.
\Cref{fig:caida-mptcp-absolute} paints the complementary picture of the dataset in
absolute numbers.
The bars (attached to the left y-axis) denote the aggregate amount of MPTCP bytes, and the line (to the right y-axis) shows the mean of MPTCP flows over four years.
We observe a $\approx$~8.6$\times$ jump in MPTCP bytes from 2016--2018 and an increase of 64\% within 2018--2019.
However, the concurrent increase in the number of MPTCP flows
hints that MPTCP is largely being used for short-lived small transfers.
Unfortunately, we cannot analyze the after-effects of MPTCP upstreaming in Linux at the beginning of 2020 from the CAIDA dataset as no trace data is available beyond 2019.
Hence we turn our attention to the MAWI traces.

\begin{figure}[t!]
	\includegraphics[width=\columnwidth]{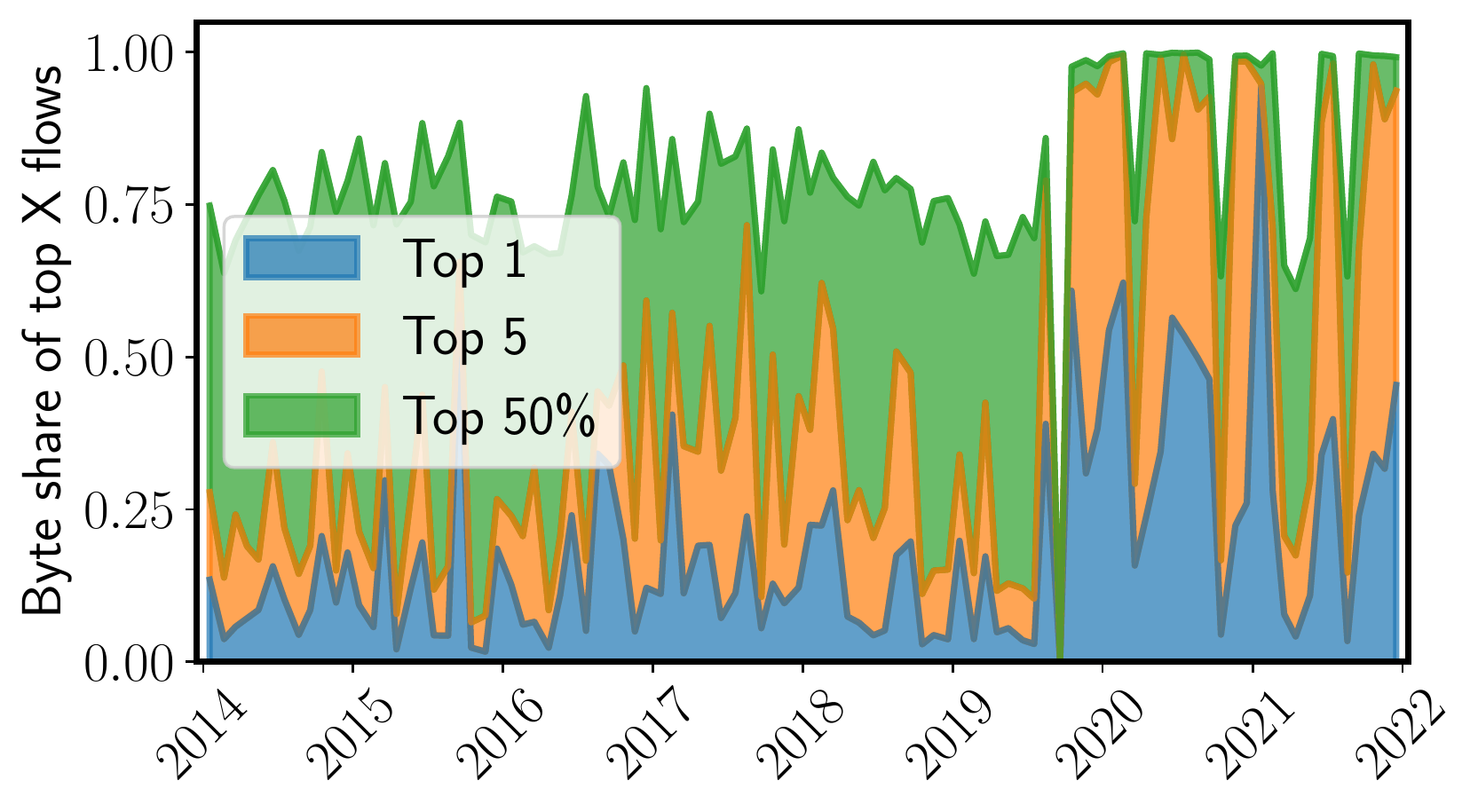}
	\caption{MPTCP flow size distribution at MAWI's samplepoint-F showing top 1, top 5, and top 50\% of flows compared to all MPTCP traffic.}
	\label{fig:mawi-mptcp-byteshare}
\end{figure}

From \Cref{fig:mawi-mptcp-share} we observe that the share of MPTCP traffic flows captured by MAWI stays relatively constant over time, making up less than 0.1\% of all TCP traffic flows.
The share of MPTCP bytes is even smaller, until the end of 2019,
as we begin to see it increase substantially, peaking at upwards of 0.4\% in June 2020 and February 2021.
Interestingly, the number of MPTCP flows remains low and does not increase.
We further investigate this phenomenon by looking at the flow size distribution over time.
To convey this distribution, we show the traffic share of the top flow, the top five flows, and the top 50\% of flows in \Cref{fig:mawi-mptcp-byteshare}.
If all flows had the same size, the green top 50\% line would be at 0.5.
Around the end of 2019, we see a drastic change in flow size distributions.
At times, a single flow makes up 50\% of all MPTCP traffic, and the top five flows make up almost all of MPTCP traffic.
This indicates that MPTCP is starting to be used and carries actual data.
We also evaluate the duration of these ``elephant flows'' and find that they last $\approx$ 30s.
The relatively short duration also explains the dips in the top 5 in 2020 and 2021, as seen in \Cref{fig:mawi-mptcp-share}.
If an elephant MPTCP flow is not present within MAWI's 15-minute daily capturing window, its distribution and traffic share drops considerably.

\begin{figure}[t!]
	\centering
	\includegraphics[width=\columnwidth]{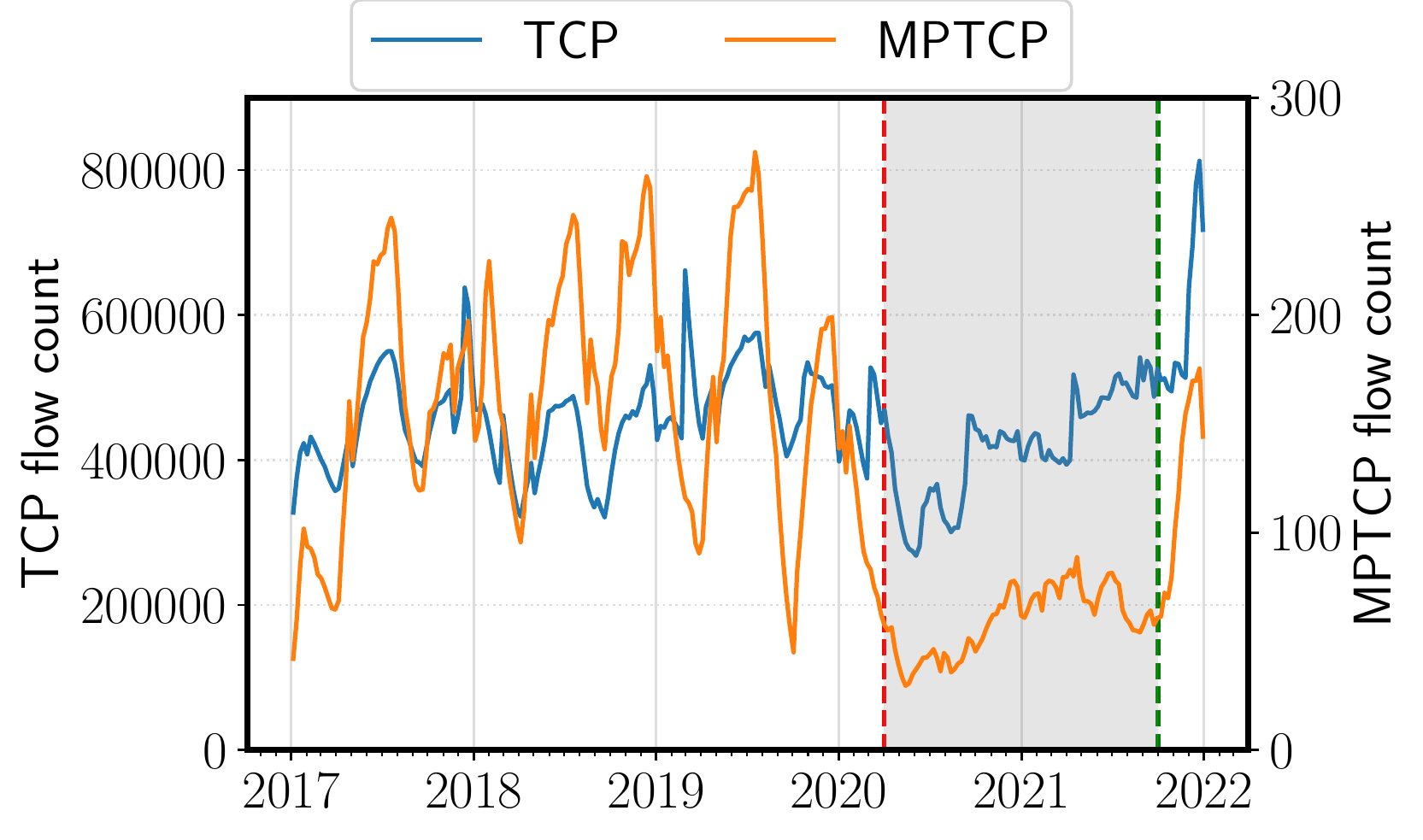}
	\caption{Exponentially weighted moving average of TCP and MPTCP flow count in MAWI captures.}
	\label{fig:mawi-smoothed-TCP-MPTCP-flowcount}
\end{figure}

To further analyze the number of flows, we process all Thursdays recorded by MAWI from 2017 until the end of 2021. \Cref{fig:mawi-smoothed-TCP-MPTCP-flowcount} shows the exponentially weighted moving average (EWMA)\footnote{\(EWMA(t)=a\times x(t) + (1-a)\times EWMA(t-1)\), with \(a=0.2\)} of the number of MPTCP and TCP flows. As can be seen, the number of TCP flows is relatively stable during the analysis period. However, the number of MPTCP flows fluctuates, and we find two peaks and valleys in each year between 2017 and 2020. These fluctuations reflect the nature of the MAWI dataset, which is an academic network capture and can be affected by the presence of students and staff on campus. At the beginning of 2020 (vertical red line), the COVID-19 pandemic leads to a switch to remote working and remote learning, which drastically impacts the TCP flows, but even more so the MPTCP flows.
The number of flows decreases, and the trend is not comparable to that of previous years. The result is not surprising, as our active scans show that mobile applications on iOS are the main MPTCP clients in the wild (see \Cref{sec:tracebox}).
At the end of 2021, many lockdown restrictions were lifted in Japan, and we see an increase in MPTCP and TCP flows.
This finding is consistent with previous work investigating the effects of the COVID-19 pandemic on Internet traffic \cite{feldmann2020lockdown,feldmann2021year,chang2021can,karamollahi2022zoomiversity,bottger2020internet,candela2021italian,lutu2020characterization}.
Furthermore, 
we also find that all captured MPTCP traffic between 2017 and 2021 in our MAWI dataset is for MPTCPv0. However, we do observe a handful of MPTCPv1 flows in 2022, all of which were to/from Apple servers (cf. \Cref{sec:discussion:sub:apple}).

\subsection{MPTCP Application Usage} \label{sec:mptcp-application-usage}

To better understand the applications used in MPTCP traffic, we map transport port numbers for MAWI and CAIDA traces to well-known port numbers used for specific services~\cite{ianaPorts}.
Additionally, we leverage Apple's list of ports used in their services to identify Apple service traffic \cite{applePorts}.
We are able to successfully map all flows to well-known ports in MAWI; except for one flow with both high ports and two flows with reserved value zero as the source port.
The latter could be attributed to misconfigured devices
\cite{maghsoudlou2021zeroing}.
For CAIDA, we find more than 80\% of source ports in the well-known range and a majority of destination ports as ephemeral, indicating that the link mainly carries server-to-client upstream traffic.

Overall, we observe six different applications utilizing MPTCP in MAWI: HTTPS, HTTP, Ident, SMB, Siri, and RDP.
However, the overwhelming majority of all MPTCP traffic is HTTPS traffic, whose lowest share is 99.5\%.
On the other hand, the application mix in the CAIDA dataset is more diverse than MAWI, as we find 15+ services using MPTCP, including HTTPS, HTTP, Spamtrap, and Microsoft services.
However, HTTPS traffic eclipses all other applications, similar to MAWI, with 99.91\% being its lowest share.
Moreover, other than very small traces of Siri in 2018, we did not discover any other instances of Apple services using MPTCP in both datasets.
Finally, using non-anonymized MAWI traces for select days in 2021, we find that all MPTCP traffic in that year is directed toward Apple servers.

\begin{tcolorbox}[title=\textit{Takeaway}, enhanced, breakable]
	The MPTCP traffic share remains consistently low over time.
	Since mid-2019, MPTCP traffic has increased steadily and now includes larger flows, indicating that MPTCP has started to see actual deployment.
	With more than 99\% of all MPTCP traffic, HTTPS over MPTCPv0 is the dominant application using MPTCP in the wild.
	We can also observe a drop in MAWI MPTCP traffic during the COVID-19 pandemic, confirming findings by related work.
\end{tcolorbox}

\section{Discussion} \label{sec:discussion}

\subsection{Case-Study: A Deeper Bite of the Apple}
\label{sec:discussion:sub:apple}

\begin{figure}[t!]
    \includegraphics[width=\columnwidth]{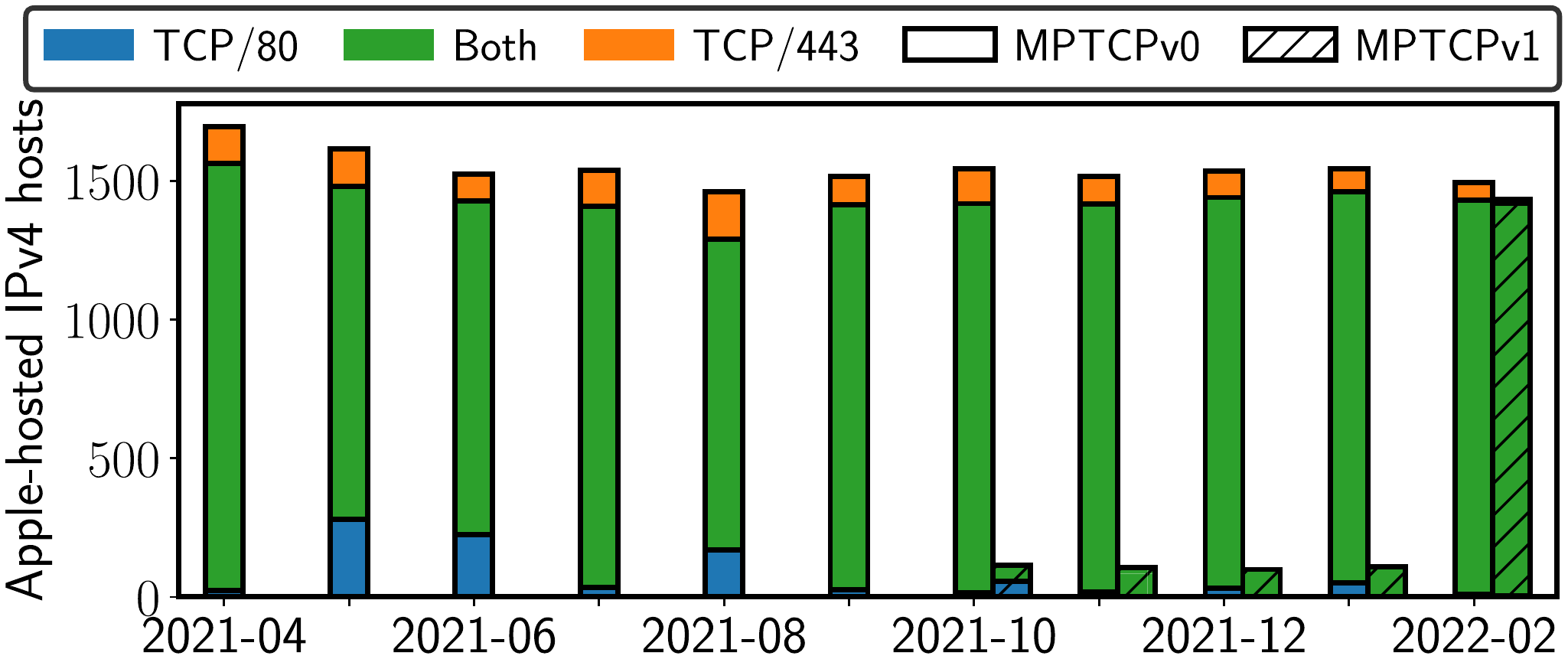}
    \caption{MPTCP IPv4 hosts within ASes owned by Apple (measurement period extended to February 2022).}
    \label{fig:ipv4-apple}
\end{figure}

Our analysis in \cref{sec:geo-distribution} shows that Apple continues to be one of the most prominent supporters of MPTCP and has a significant deployment of servers that support MPTCPv0 and MPTCPv1 for services on port 80 and port 443 in IPv4 and IPv6.
We also show through our MAWI traffic analysis in \cref{sec:mptcpTraffic} that Apple also actively utilizes MPTCP to enhance the performance of its system services, e.g., Siri.
As such, we now provide a deeper look into the MPTCP support within Apple, explicitly understanding how Apple is integrating the newly standardized MPTCPv1 within its infrastructure.

\Cref{fig:ipv4-apple} shows the IPv4 hosts within ASes owned by Apple (i.e., ASN6185 \& ASN714) that reportedly support MPTCPv0 and MPTCPv1 in our ZMap scans. 
Since we started scanning for MPTCPv1 in our ZMap scans in March 2021, we analyze the period beginning in April 2021 and extend our end date to February 2022 (totaling \emph{ten} months).
Firstly, we find that Apple's support over MPTCPv0 remains relatively consistent throughout our analysis period, with almost 1500+ hosts supporting MPTCPv0 in our ZMap scans.
We also observe that a large majority of these hosts (exceeding 90\%) provide services over both port 80 and port 443---further supporting our port breakup analysis in \Cref{fig:ipv4-port-breakup}.
Interestingly, the plot shows that despite MPTCPv1 being available in the default Linux kernel 5.6 since March 2020~\cite{upstream_linux}, Apple did not utilize MPTCPv1 for its services until October 2021, i.e., almost 20 months after its availability.
However, MPTCPv1 support within Apple remained limited compared to MPTCPv0 ($\lesssim$ 100 MPTCPv1 hosts), which remained consistent until the end of January 2022.   
In February 2022, we observe a significant uptick in MPTCPv1 support within Apple, increasing almost 11$\times$ ($\approx$ \emph{1400} hosts), approximately equaling MPTCPv0 support in the same month.
Similarly, in IPv6, we find that Apple had no MPTCPv1 deployment until September 2021, with $\approx$ 1200 MPTCPv0 IPs.
In October 2021, Apple added 90 dual-protocol MPTCPv1 IPs, with the bulk of 1100 additional IPs being added in February 2022.

\begin{figure}[t!]
    \includegraphics[width=\columnwidth]{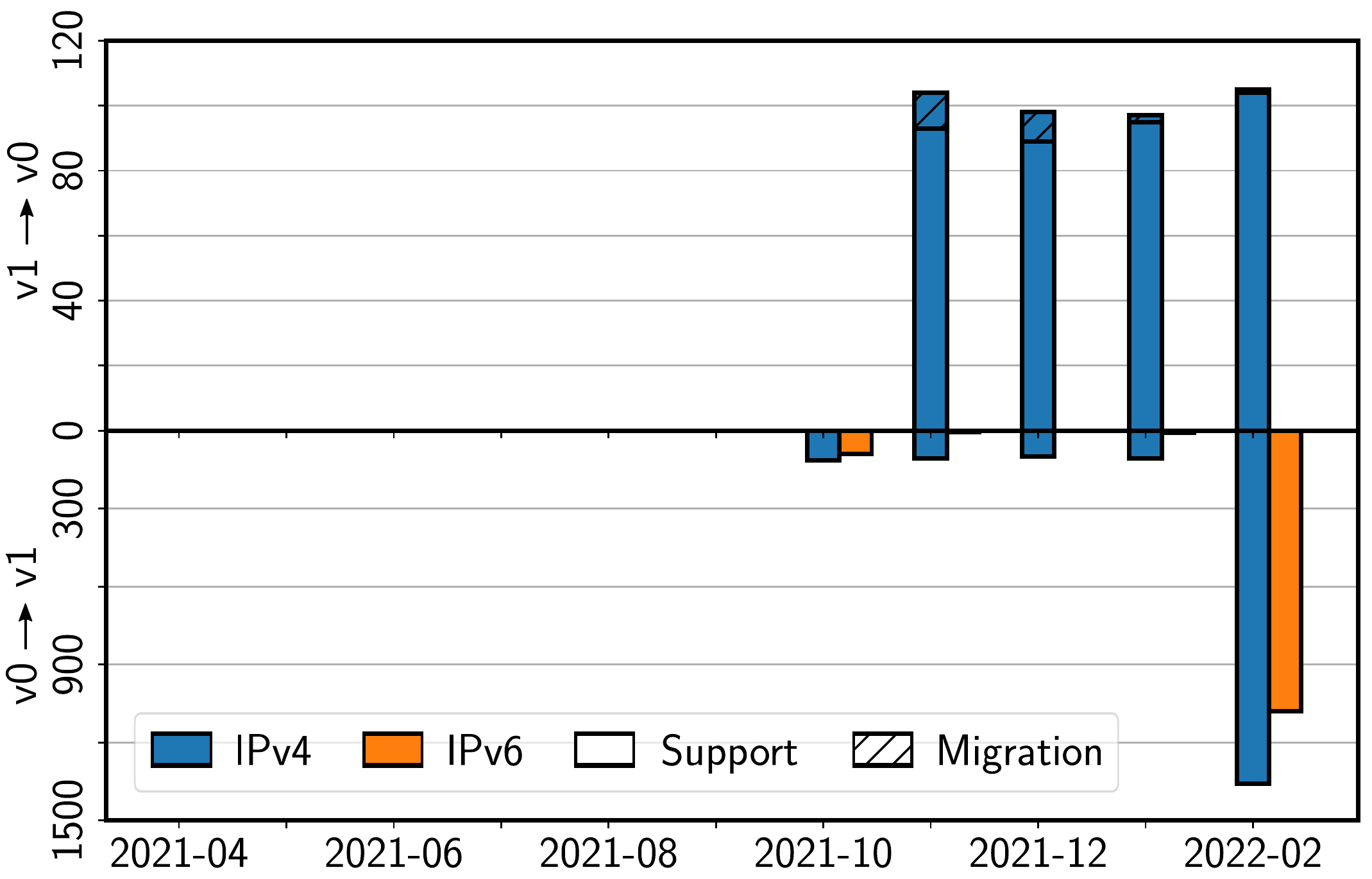}
    \caption{Added support and migration between MPTCPv0 and MPTCPv1 within Apple ASes (\emph{note the different y-axis scale for both versions}). The measurement period extends to February 2022.}
    \label{fig:apple-migration}
\end{figure}

The sudden support for MPTCPv1 within Apple is quite peculiar and can be due to multiple reasons.
For one, Apple may be migrating its existing MPTCPv0 infrastructure to MPTCPv1.
The simultaneous support for both versions, however, makes this assumption unlikely.
Similarly, Apple might be employing both MPTCPv0 and MPTCPv1 for different (non-overlapping) services.
We investigate the reason behind the event further by analyzing the support migration in consecutive months between the two MPTCP versions within Apple infrastructure (see \Cref{fig:apple-migration}).
The bottom half of the figure denotes hosts that supported MPTCPv0 on either TCP/80 or TCP/443 in the previous month and now added support for MPTCPv1.
Similarly, the top half of the figure denotes hosts that supported MPTCPv1 in the previous month and now support MPTCPv0.
The dashed region denotes ``migrated'' hosts, \ie hosts that no longer support the previously supported MPTCP version.
On the other hand, the ``support'' numbers denote hosts that did not support one of the two versions in the previous months but now show support for both MPTCPv0 and MPTCPv1.
The plot shows a direct correlation with \Cref{fig:ipv4-apple}. 
We find that the minor uptick in MPTCPv1 support within Apple in October 2021 over IPv4 and IPv6 is due to MPTCPv0 hosts that also started supporting MPTCPv1.
In the following few months, we observed that many IPv4 hosts that supported either MPTCPv0 or MPTCPv1 in the previous month began supporting the other version.
Note that the trend is largely missing from IPv6.
Interestingly, we find that only 11 IPv4 hosts migrated their support from MPTCPv1 to MPTCPv0 (we did not find any migration in the opposite direction or within IPv6).
On the other hand, the result shows that the significant uptick in MPTCPv1 support within Apple in February 2022 is due to IPv4 and IPv6 hosts that previously supported MPTCPv0 also starting to support MPTCPv1 ($\approx$96\% of total MPTCPv1 hosts this month also support MPTCPv0).

\subsection{Factors Affecting Future MPTCP Deployment}
In this work, we showcase that MPTCP's adoption and support globally have been slow but steady. 
Originating as a research project with ambitions to maximize throughput within datacenter environments~\cite{mptcp-datacenters}, MPTCP has evolved significantly and is now primarily used in commercial (mobile-focused) networks---thanks to its capabilities to multiplex over multiple heterogeneous paths. 
However, we observe that the rise in \emph{infrastructure} size is not yet reflected in MPTCP's \emph{traffic share} compared to TCP in the Internet (see \Cref{sec:mptcpTraffic}).
We also find that MPTCPv0 is highly susceptible to middleboxes in the Internet -- however, this shortcoming seems to have been successfully plugged by the introduction of MPTCPv1.
This makes the potential future for MPTCP ``somewhat attractive'' to the Internet community, provided the state of the transport protocol standards remains unchanged.
For instance, as we highlight in \Cref{sec:related}, there is a significant interest in the community to support and utilize QUIC for supporting service operations in the Internet.
Thanks to its flexibility, the protocol allows operators to experiment with inherent functionality to best fit the needs of different use-cases~\cite{beyond-quic}.
Additionally, being end-to-end encrypted, QUIC remains largely unaffected by middleboxes in the Internet (unlike TCP).
As a result, several researchers are actively working on providing similar multipath capabilities to QUIC.
As of now, there are three different multipath proposals for QUIC: \cite{draft-connick}, \cite{draft-huitema} and \cite{draft-liu} -- each proposing different features to support multipath.
While no specific multipath over QUIC proposal has yet been selected for standardization, thanks to the inherent disagreements on core design~\cite{de2021packet}, the authors of all three proposals wrote a common draft~\cite{draft-lmbdhk} which has a high likelihood of being standardized.
Considering the popularity of QUIC, the potential standardization of MP-QUIC may threaten the deployment and continued usage of MPTCP.
As such, we envision the widespread adoption of MPTCP to be possible only if it is embraced by service-providing organizations, in addition to Apple, for supporting real applications in the Internet.

\subsection{Limitations of our Methodology}

Our methodology does not capture client-side MPTCP deployments, including MPTCP proxy solutions that work only when the client establishes an MPTCP connection. The passive data analysis (see \Cref{sec:mptcpTraffic}) only focuses on MPTCP support and not MPTCP usage. 
We plan to plug these limitations in a future study, with MPTCPv1 results being already available at \texttt{\url{https://mptcp.io}}.

\section{Conclusion}\label{sec:conclusion}

This paper presented the first broad multi-faceted assessment on MPTCPv0 and MPTCPv1 deployment in the Internet.
We studied both the \textit{infrastructure}, by probing the entire IPv4 address space and an IPv6 hitlist for MPTCP-capable IPs, and \textit{traffic share} at two geographically diverse vantage points.
We identified middleboxes that impact both MPTCP scanning attempts and user traffic during the course of our study, hence providing the most accurate picture of \emph{true} MPTCP deployment to date.
We observed a steady growth in MPTCP-enabled IPs that support HTTP and HTTPS in our 18-month investigation period, reaching $\approx$ 13k+ and 1k in December 2021 for IPv4 and IPv6, respectively.
The growth is primarily driven by Apple and ISPs across the globe that rely on the protocol to enhance their services.
\section{Acknowlegment}\label{sec:ack}
We are grateful to Olivier Bonaventure for his valuable feedback and comments on this paper. We would like to thank the network operators at the Chair of Connected Mobility and Technical University of Munich, Germany, especially Simon Zelenski, for supporting our scans and managing the abuse report emails. We also thank Kenjiro Cho for providing access to the MAWI traces.
This research is funded by EU Celtic project ``Piccolo'' (C2019/2-2), Bavarian Ministry of Economic Affairs project ``6G Future Lab'' and the German Federal Ministry of Education project ``6G-Life''. 
 
\balance
\bibliographystyle{IEEEtran}
\bibliography{bibliography.bib}
\end{document}